\documentclass[a4paper,fleqn,usenatbib]{mnras}
\usepackage{newtxtext,newtxmath}

\usepackage[T1]{fontenc}
\usepackage{ae,aecompl}

\usepackage[utf8]{inputenc}
\usepackage{etoolbox}
\makeatletter
\patchcmd\@combinedblfloats{\box\@outputbox}{\unvbox\@outputbox}{}{%
   \errmessage{\noexpand\@combinedblfloats could not be patched}%
}%
 \makeatother
\usepackage{graphicx}	
\usepackage{amsmath}	
\usepackage{amssymb}	
\usepackage{xcolor}
\usepackage{hyperref}
\usepackage{cleveref}

\newcommand{\kms}{\,km\,s$^{-1}$}
\newcommand{\5}{[O III] $\lambda$5007}
\newcommand{\oiii}{[O III] $\lambda \lambda$4959, 5007}
\newcommand{\nii}{[N II] $\lambda \lambda$6548, 6584}
\newcommand{\sii}{[S II] $\lambda \lambda$6716, 6731}

\title[The multi-wavelength gas kinematics of GMASS 0953]{A multi-wavelength study of a massive, active galaxy at $z\sim 2$: coupling the kinematics of the ionized and molecular gas}
\author[F. Loiacono et al.]{
Federica Loiacono$^{1, 2}$\thanks{E-mail: federica.loiacono2@unibo.it},
Margherita Talia$^{1, 2}$,
Filippo Fraternali$^{3, 1}$,
Andrea Cimatti$^{1, 4}$,
\newauthor \ Enrico M. Di Teodoro$^5$ and Gabriel B. Caminha$^{3}$
\\
$^{1}$Dipartimento di Fisica e Astronomia, Universit$\grave{a}$ di Bologna, via Gobetti 93/2, I-40129 Bologna, Italy\\
$^{2}$INAF-Osservatorio di Astrofisica e Scienza dello Spazio, via Gobetti 93/3, I-40129 Bologna, Italy\\
$^{3}$Kapteyn Astronomical Institute, University of Groningen, Postbus 800, 9700 AV, Groningen, The Netherlands\\
$^{4}$INAF-Osservatorio Astrofisico di Arcetri, Largo E. Fermi 5, 50125 Firenze, Italy\\
$^5$Research School of Astronomy and Astrophysics - The Australian National University, Canberra, ACT, 2611, Australia} 

\date{Accepted XXX. Received YYY; in original form ZZZ}

\pubyear{2019}

\begin{document}
\label{firstpage}
\pagerange{\pageref{firstpage}--\pageref{lastpage}}
\maketitle

\begin{abstract}
We report a multi-wavelength study of the massive ($M_{\star} \gtrsim 10^{11} \rm{M}_{\odot}$), $z\sim2$ star-forming galaxy GMASS 0953, which hosts an obscured AGN.
We combined near-infrared observations of the GNIRS, SINFONI and KMOS spectrographs
to study the kinematics of the \5 and H$\alpha$ emission lines.
Our analysis shows that GMASS 0953 may host an ionized disc extending up to 13 kpc, which rotates at a velocity of $V_{\rm{ion}} = 203^{+17}_{-20}$ \kms\ at the outermost radius. Evidence of rotation on a smaller scale ($R \sim 1$ kpc) arises from the CO(J=6-5) line. The central velocity $V_{\rm{CO}} = 320^{+ 92}_{-53}$ \kms\ traced by the molecular gas is higher than $V_{\rm{ion}}$, suggesting that the galaxy harbors a multi-phase disc with a rotation curve that peaks in the very central regions. The galaxy appears well located on the $z = 0$ baryonic Tully-Fisher relation. We also discuss the possibility that the \5 and H$\alpha$ velocity gradients are due to a galactic-scale wind.
Besides, we found evidence of an AGN-driven outflow traced by a broad blueshifted wing affecting the \5 line, which presents a velocity offset $\Delta v = -535 \pm 152$ \kms\ from the systemic velocity. 
Because of the short depletion timescale ($\tau_{\rm{dep}}\sim 10^8$ yr) due to gas ejection and gas consumption by star formation activity, GMASS 0953 may likely evolve into a passive galaxy. However, the role of the AGN in depleting the gas reservoir of the galaxy is quite unclear because of the uncertainties affecting the outflow rate.
\end{abstract}

\begin{keywords}
galaxies: active -- galaxy: evolution -- galaxies: kinematics and dynamics -- galaxies: high-redshift
\end{keywords}



\section{Introduction}
In the last years significant progress has been made in the study of $z\sim 2$ massive ($M_{\star} \gtrsim 10^{11}$ M$_{\odot}$) galaxies. 
Investigating these systems is pivotal to shed light on the mechanisms that regulate the mass assembly at the peak epoch of both star formation and nuclear activity (e.g. \citealt{2014ARA&A..52..415M}).
Near-infrared (NIR) observations using integral field spectroscopy have revealed that rotating discs are common in these objects \citep{2009ApJ...706.1364F,2015ApJ...799..209W}, with important implications for their growth mode.
The interplay between active galactic nuclei (AGN) and their hosts have been deeply investigated through spatially resolved observations of ionized outflows traced by strong emission lines (e.g. \citealt{2012MNRAS.426.1073H, 2014ApJ...796....7G, 2015A&A...580A.102C}) while at the sub-millimeter wavelengths interferometers as the NOrthern Extended Millimeter Array (NOEMA) and the Atacama Large Millimeter Array (ALMA) have provided key information on the molecular gas content of massive systems through CO lines and dust emission (e.g. \citealt{ 2016ApJ...820...83S, 2018ApJ...853..179T}).\\
\indent Nevertheless, a number of questions remain open. A first issue concerns the so-called "quenching" of star formation in massive spheroids. Several studies show that massive spheroidal galaxies are dominated by old stellar populations already at $1 < z < 2$ 
(e.g. \citealt{2004Natur.430..181G, 2004Natur.430..184C, 2008A&A...482...21C, 2015ApJ...808..161O}). Because of their size, considerably smaller compared to present-day ellipticals, and high stellar density the progenitors of these systems are thought to be compact star-forming galaxies (SFGs) at $1.5 < z < 3$ (e.g. \citealt{2015ApJ...813...23V}). These objects are often characterized by a short depletion timescale inferred from the molecular gas content \citep{2016ApJ...832...19S, 2017ApJ...851L..40B, 2017ApJ...841L..25T} suggesting that they are rapidly exhausting their internal reservoir for star formation. However, new gas can be accreted from the circumgalactic and intergalactic medium (CGM/IGM) as shown by numerical simulations (\citealt{2005MNRAS.363....2K, 2009Natur.457..451D}), which may keep sustaining the star formation activity.
To explain the quenching of star formation in spheroidal galaxies is thus necessary a mechanism that removes or heats the accreted gas, in order to make it not available to form stars. 
AGN could be crucial in this context through feedback mechanism (see \citealt{2012ARA&A..50..455F} for a review). 
In particular, high-velocity winds ($v>$ 500 \kms) extending on kpc-scales driven by the AGN may be efficient in depleting the galaxy of the accreted gas, cutting off the fuel for star formation. 
This ejective feedback is implemented in models as a way to prevent both black hole and galaxy growth (e.g. \citealt{2005ApJ...635L.121K, 2005ApJ...620L..79S, 2005Natur.433..604D, 2016MNRAS.458..816H, 2017MNRAS.465.3291W, 2018MNRAS.475..624N}).
However, if on the theoretical side AGN feedback is commonly invoked to explain the observed properties of galaxies, from the observational point of view we still lack a robust understanding of how AGN shape their hosts (\citealt{2018NatAs...2..179C}).\\
\indent An important clue about the influence of AGN feedback in turning massive SFGs to passive spheroids may arise from the population of the red sequence in the color-mass plane \citep{2008A&A...483L..39C} by spheroidal galaxies between $z\sim2.4$ and $z\sim1.4$ and the decrease of AGN activity in SFGs between the same redshifts \citep{2013ApJ...779L..13C}. Rest-frame UV, stacked spectra of active galaxies in the sample of \citet{2013ApJ...779L..13C} show absorption lines with average velocity offset $\Delta v\gtrsim - 500$\kms\ that are signatures of gas outflows, highlighting that the progenitors of passive spheroids may be found among the SFGs with AGN feedback in action.
However, the picture is still hazy and detailed observations of systems hosting AGN-driven outflows are needed especially at $z\gtrsim 2$, i.e. the formation epoch of the spheroids progenitors, in order to assess the efficiency of AGN winds in suppressing star formation.\par
\indent Another key issue concerning $z\sim2$ massive galaxies relates to the kinematics of the insterstellar medium (ISM). 
Using rotation curves to investigate the kinematics of $z\sim 2$ discs provides important constraints on the dark matter content and on the evolution of scaling relations such as the Tully-Fisher \citep{1977A&A....54..661T} whose matching is an important test bench for theoretical models (e.g. \citealt{2017MNRAS.464.4736F}).
Most of the studies at $z\sim 2$ uses the ionized phase of the ISM to trace the kinematics of massive galaxies, which has been deeply investigated through large programs such as SINS \citep{2009ApJ...706.1364F} and KMOS$^{\rm{3D}}$ \citep{2015ApJ...799..209W}. 
The ionized gas has though the drawback to be more affected compared to the cold gas by non-circular motions (e.g. outflows), which may make this tracer, if considered alone, not representative of the overall kinematics (e.g. \citealt{2018MNRAS.tmp.1721L}).
A multi-wavelength approach that employs different ISM tracers (e.g. molecular, ionized), is hence required in order to build a robust picture of the kinematics of $z\sim 2$ objects.
At the moment this kind of studies is rare, regarding just a scanty number of high-$z$ galaxies (see for example \citealt{2013ApJ...773...68G, 2017ApJ...846..108C, 2019ApJ...871...37H, 2018MNRAS.tmp.1721L, 2018ApJ...854L..24U}) because of the lack of multi-wavelength spectroscopic data at high redshift.\\
\begin{figure}
\begin{center}
\includegraphics[scale=0.3]{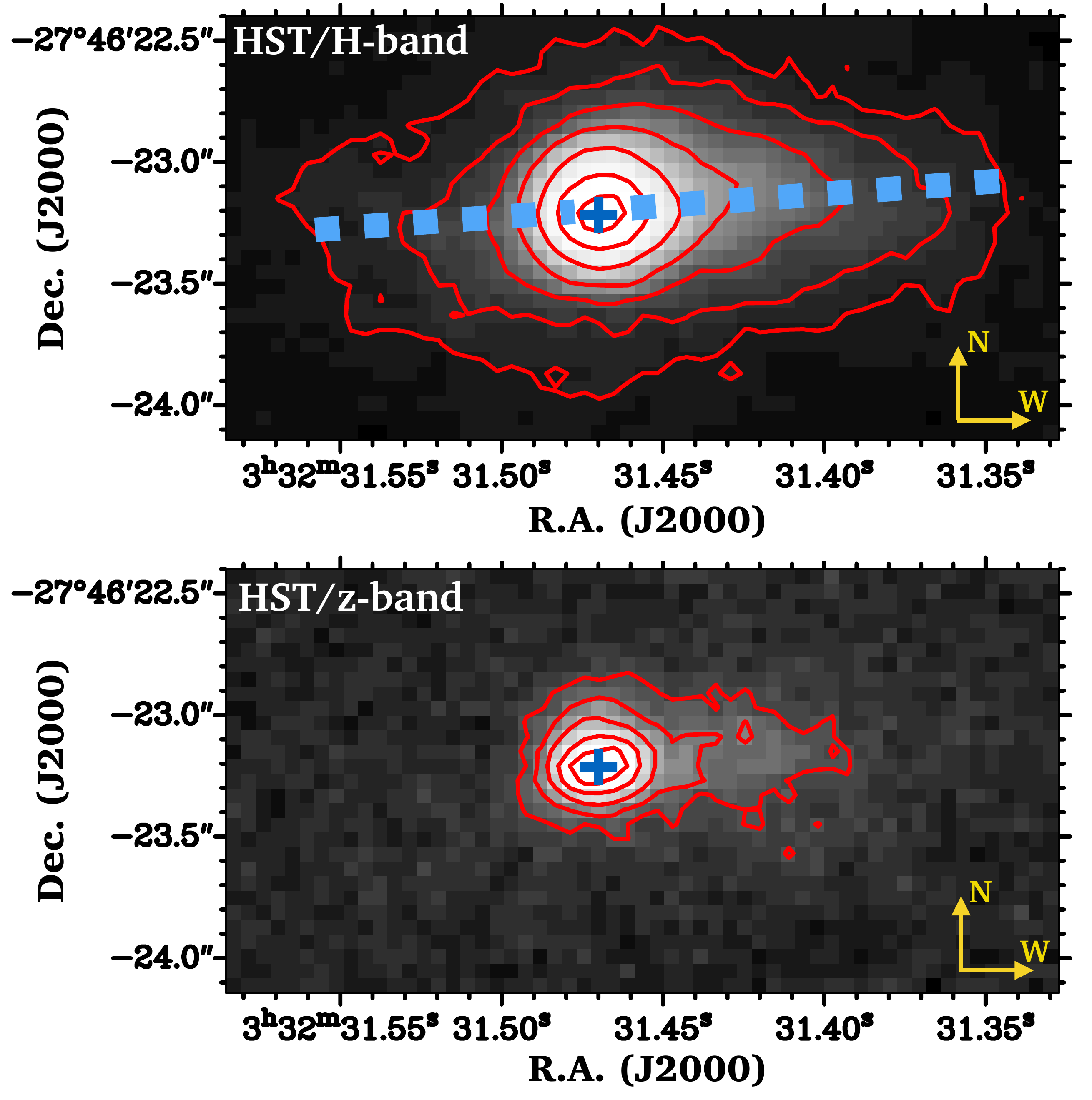}
\end{center}
\caption{\textit{H}-band (1.6 $\mu$m, upper panel) and  \textit{z}-band (0.85 $\mu$m, bottom panel) images of GMASS 0953 (\textit{HST}/WFC3 and ACS). The lowest contour is at 5$\sigma$. The blue crosses show the emission peak, which we take as the photometric and dynamical center of the galaxy. The central region appears extremely bright and compact with an effective radius $R_e \sim 2.5$ kpc, estimated from the \textit{H}-band image \citep{2014ApJ...788...28V}. The dashed line indicates the major axis (position angle $\phi=94$\degr).}
\label{fig:hst_h}
\end{figure}
\\
\indent Here we report a multi-wavelength study of the $z\simeq 2.226$ galaxy GMASS 0953. This is a massive, compact SFG, which hosts an obscured AGN. 
This source has been already studied by \citealt{2018MNRAS.476.3956T}, who investigated its molecular content revealing the existence of a rotating disc traced by the CO(J=6-5) line. Moreover, \citealt{2017A&A...602A..11P} found a very short depletion timescale for the molecular gas, which implies that the galaxy is rapidly consuming its gas reservoir and it will likely evolve into a passive system. 
The present work integrates the aforementioned picture adding information on the ionized gas kinematics. In particular, we discuss how the ionized gas motions relate to the molecular phase and what is the effect of the AGN feedback on the galaxy.
In our analysis, we used NIR observations of \5 and H$\alpha$ combining a slit spectrum, obtained with the Gemini Near InfraRed Spectrograph (GNIRS), with integral field data collected by the Spectrograph for INtegral Field Observations in the Near Infrared (SINFONI) and the K-band Multi-Object Spectrograph (KMOS). 
The use of different instruments that work at similar wavelengths has the advantage to corroborate the results arising from independent measurements and provides also complementary details (see section~\ref{sec:data}).
The paper is organized as follows: in section~\ref{sec:portrait} we describe the main properties of GMASS 0953 while section~\ref{sec:data} concerns the data analysis. The multi-wavelength kinematics is discussed in section~\ref{sec:kine} and section~\ref{sub:multiDisc}, while we study the AGN feedback in section~\ref{sec:feedback}. 
We finally summarize the obtained results in section~\ref{sec:conclusions}. 
We adopt a $\Lambda$CDM cosmology with $\Omega_{\Lambda} = 0.71$, $\Omega_{\rm{M}} = 0.29$ and  $H_0 = 69.6$ \kms Mpc$^{-1}$. In this cosmology 1$''$ corresponds to 8.4 kpc at $z\simeq 2.226$.
We assume a \citet{2003PASP..115..763C} IMF.
\section{GMASS 0953: a global portrait}
\label{sec:portrait}
\begin{table*}
	\centering
	\caption{Main properties of GMASS 0953}
\resizebox*{0.98\textwidth}{!}{	
	\begin{tabular}{cccccccccccc} 
		\hline \hline
		R.A. & Dec. & $z$ & $R_{\rm{eff},H\rm{-band}}$ &$M_{\star}$ & SFR$_{\rm{IR}}$ & SFR$_{\rm{H}\alpha}$ &$L_X$(2-10 keV) & HR&$N_{\rm{H}} $&E(B-V)&$n_\text{e}$\\
        (J2000)&(J2000)&&(kpc)&$(10^{11}\rm{M}_{\odot})$&(M$_{\odot}$ yr$^{-1}$)&(M$_{\odot}$ yr$^{-1}$)&(erg s$^{-1}$)&&($10^{24}$cm$^{-2}$)&&(cm$^{-3}$)\\
        (1)&(2)&(3)&(4)&(5)&(6)&(7)&(8)&(9)&(10)&(11)&(12)\\
        \hline
        	3:32:31.4&-27:46:23.2& 2.226&2.5&$1.15 \pm 0.10$&$214 \pm 20$&498 $\pm$ 33&$5\times 10^{44}$&0.6&$ 4.4^{+4.7}_{-1.7}$&$ 0.8 \pm 0.3$&$\sim 500$\\
        	\hline
		\hline
\multicolumn{12}{p{.98\textwidth}}{\textbf{Notes.} Column description: (1), (2) sky-coordinates of GMASS 0953 in the \textit{HST}/WFC3 \textit{H}-band image. (3) Redshift derived from the \5 narrow line. (4) Effective radius measured on the \textit{HST}/WFC3 \textit{H}-band image \citep{2014ApJ...788...28V}. (5) Stellar mass derived from SED fitting. (6), (7) Star formation rate derived from the infrared luminosity and from H$\alpha$. (8) X-ray intrinsic luminosity (Dalla Mura et al., in prep.). (9) Hardness ratio \citep{2017ApJS..228....2L}. (10) Column density (Dalla Mura et al., in prep.). (11) Color excess derived from the H$\alpha$ and H$\beta$ line ratio. (12) Electron number density evaluated from the \sii\ line ratio.}
	\end{tabular}}
	\label{tab:info}
\end{table*}
GMASS 0953 (also known as K20-ID5, GS3-19791, 3D-HST GS30274, e.g. \citealt{2004ApJ...600L.127D,2017A&A...602A..11P, 2018ApJ...855...97W}) is a massive SFG at $z\simeq 2.226$ located in the Chandra Deep Field South (CDFS; \citealt{2002ApJS..139..369G}) (see Table~\ref{tab:info} for the main properties of the galaxy). 
It presents an irregular morphology (Figure~\ref{fig:hst_h}), visible in both the \textit{HST}/WFC3 (\textit{H}-band, 1.5 $\mu m$) and \textit{HST}/ACS (\textit{z}-band, 0.85 $\mu$m) images, which appears bright and compact in the central region ($R_{\rm{eff}, H\rm{-band}} \sim 2.5$ kpc, corresponding to 0.3$''$; \citealt{2014ApJ...788...28V}), showing also an off-nuclear emission to the west, interpreted as an asymmetric stellar disc or a merger remnant \citep{2018ApJ...855...97W, 2015ApJ...813...23V}.\\
\indent The galaxy was selected among the AGN-hosts of \citet{2013ApJ...779L..13C}, which represent a subsample of the GMASS survey \citep{2013A&A...549A..63K}, including active galaxies with a stellar mass $M_{\star} > 10^{10}$ M$_{\odot}$ and a X-ray luminosity $L_X > 10^{42.3}$ erg s$^{-1}$. The AGN activity of GMASS 0953 was inferred from the X-ray luminosity $L_X\sim 5\times 10^{44}$ erg s$^{-1}$ corrected for obscuration (Dalla Mura et al., in prep.) and the hardness ratio\footnote{The hardness ratio is defined as HR = (H - S)/(H + S) where H and S are the photon counts in the hard (2-10 keV) and soft (0.5-2 keV) X-ray band respectively.} HR = 0.6 \citep{2017ApJS..228....2L}, both higher than those measured in purely star-forming systems \citep{2004ApJ...607..721N}. The optical emission line ratios are also consistent with an AGN \citep{2014ApJ...781...21N} even if shock ionization by a strong galactic wind is also possible \citep{2005ApJ...622L..13V}. Indications of outflowing material, likely boosted by the AGN, were observed in the rest-frame optical spectrum \citep{2014ApJ...787...38F, 2014ApJ...796....7G}, in the UV absorption lines \citep{2013ApJ...779L..13C} and we found further evidence in this work.
On the other hand, the AGN activity does not emerge from the UV emission lines. This was verified by the measurement of the line ratios of C IV $\lambda$1550, He II $\lambda$1640 and C III] $\lambda$1909 using MUSE data (P. Rosati, private communication). These ratios place GMASS 0953 closer to the observed SFGs than to the AGN in the diagnostic diagram of \citet{2016MNRAS.456.3354F}, likely because of the high obscuration ($N_H= 4.4^{+4.7}_{-1.7}\times 10^{24}$ cm$^{-2}$, Dalla Mura et al., in prep.).\\
\indent Thanks to the available photometric spectral energy distribution (SED\footnote{The photometric SED of GMASS 0953 was obtained collecting data from MUSIC \citep{2006A&A...449..951G}, SPITZER/MIPS \citep{2011A&A...528A..35M}, Herschel/PACS \citep{2013A&A...553A.132M} and SPIRE \citep{2010MNRAS.409...48R}, ALMA \citep{2017A&A...602A..11P, 2018MNRAS.476.3956T, 2018ApJ...853...24U}.}) from the UV to sub-mm wavelengths, we estimated several properties of the galaxy.
The stellar mass $M_{\star}  = (1.15 \pm 0.10) \times 10^{11}$ M$_{\odot}$ was evaluated from the SED decomposition applying a modified MAGPHYS code \citep{2008MNRAS.388.1595D} that includes the AGN component, relevant in the mid-infrared \citep{2013A&A...551A.100B}, as done by  \citet{2014MNRAS.439.2736D}. The derived infrared luminosity $L_{\rm{IR}}$ in the rest-frame 8 - 1000 $\mu$m, corrected for the AGN contribution, was used to estimate the star formation rate SFR$_{\rm{IR}}$ $ = 214 \pm 20$ M$_{\odot}$yr$^{-1}$  through the \citet{1998ARA&A..36..189K} relation, rescaled to a \citet{2003PASP..115..763C} IMF.
The stellar mass and the SFR place GMASS 0953 on the main sequence at $z\sim 2$ (e.g. \citealt{2007ApJ...670..156D}).\\
\indent The kinematical properties of GMASS 0953 were investigated by \citet{2009ApJ...706.1364F}, who inferred an upper limit for the dynamical mass $M_{\text{dyn}} < 5.8 \times 10^{11}$ M$_{\odot}$. 
\citet{2018ApJ...855...97W} modelled the H$\alpha$ kinematics revealing a disky structure extending up to 13 kpc with an average rotation velocity of 200 \kms. Evidence of rotation on a smaller scale was found by \citet{2018MNRAS.476.3956T} studying the CO (J=6-5) transition produced by a rapidly rotating ($V_{\rm{CO}} =$ 320$^{+ 92}_{-53}$ \kms) and very compact core ($R_{\text{CO}} = 0.75 \pm 0.25$ kpc). \citet{2018MNRAS.476.3956T} found also a short depletion timescale for the molecular gas ($\tau_{\rm{dep}} < 150$ Myr) and a low molecular gas fraction ($f_g < 0.2$) in agreement with \citet{2017A&A...602A..11P}, which imply that GMASS 0953 might quickly cease its star formation activity, likely turning into a passive galaxy.
\section{Data analysis}
\label{sec:data}
In the sections below we describe the spectroscopic dataset and the line fitting methods that we used to evaluate the physical properties of GMASS 0953. We also discuss the derivation of total flux maps from the integral field data, which show the spatially resolved emission line morphology. 
We focus in particular on the \5 and H$\alpha$ lines that are then used to model the kinematics (see section~\ref{sec:kine})
\subsection{GNIRS one-dimensional spectrum}
\label{sub:gnirs}
\begin{figure*}
\begin{center}
\includegraphics[scale=0.62]{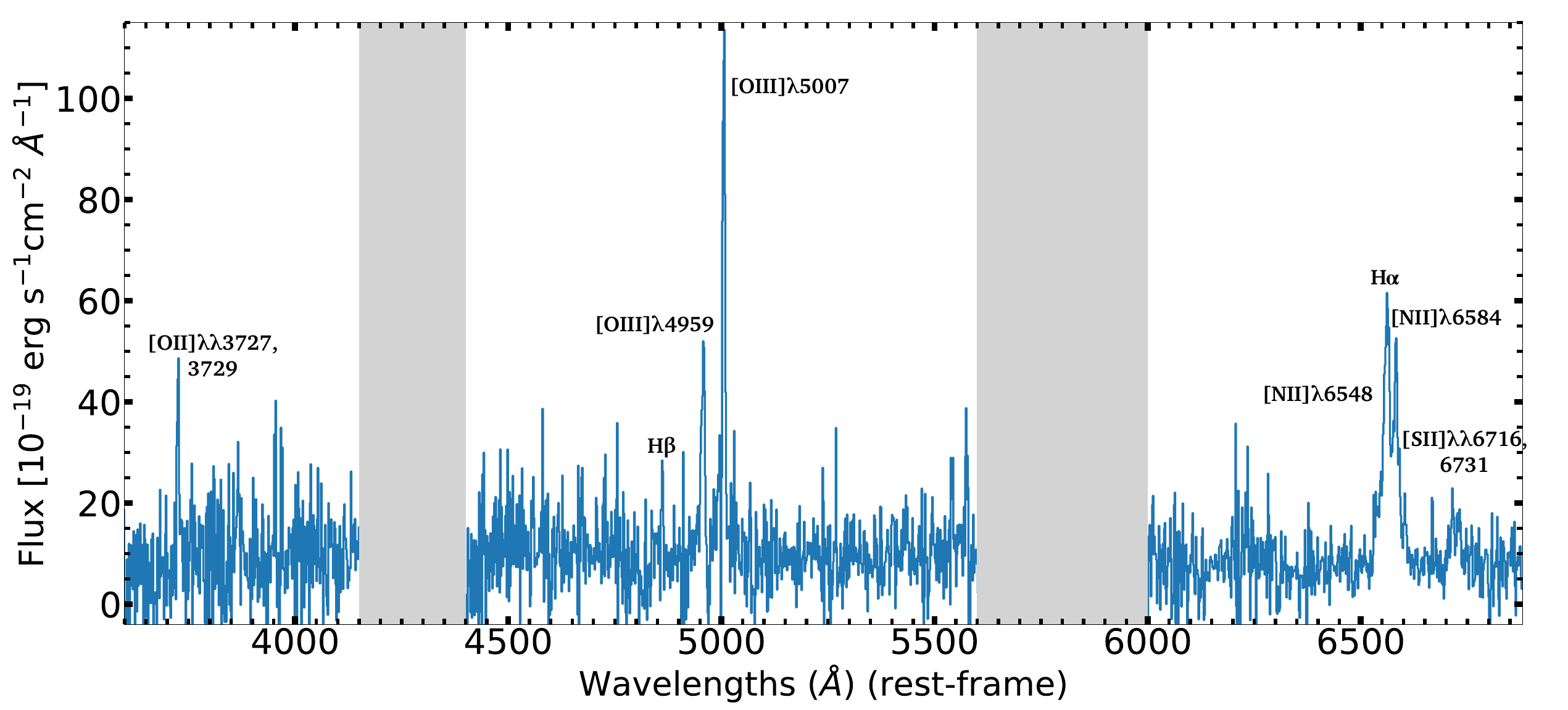}
\includegraphics[scale=0.5]{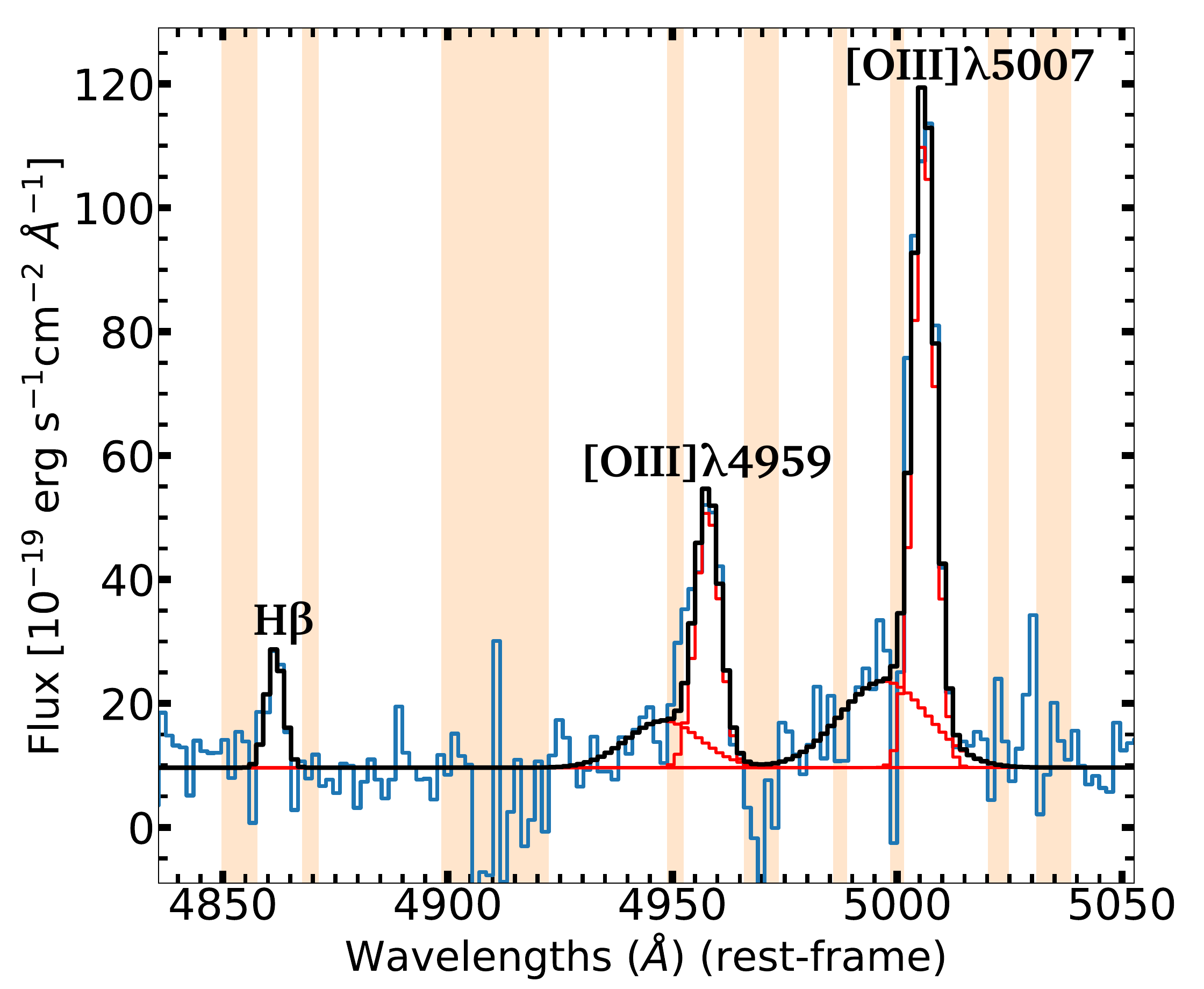}
\includegraphics[scale=0.3]{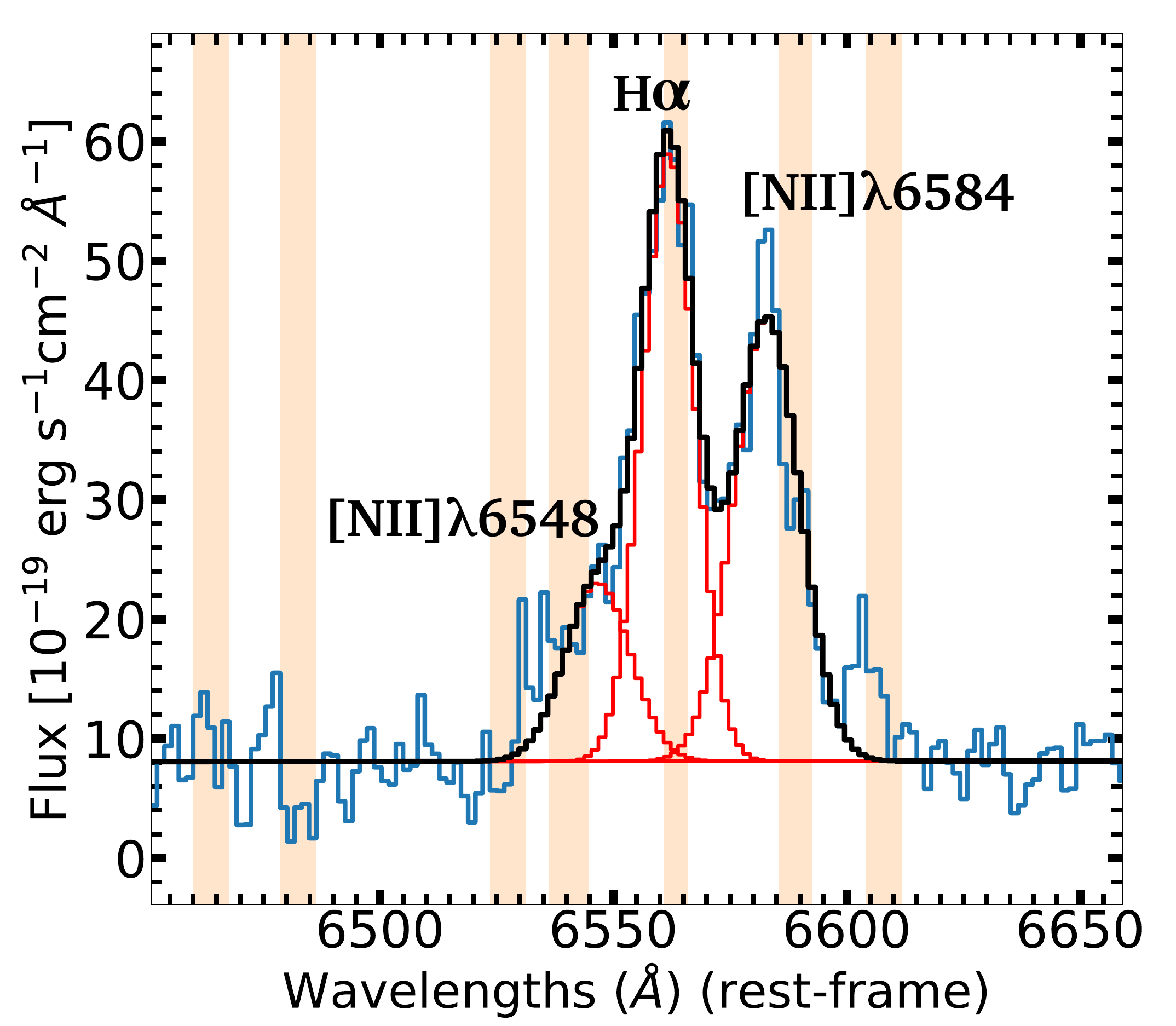}
\includegraphics[scale=0.3]{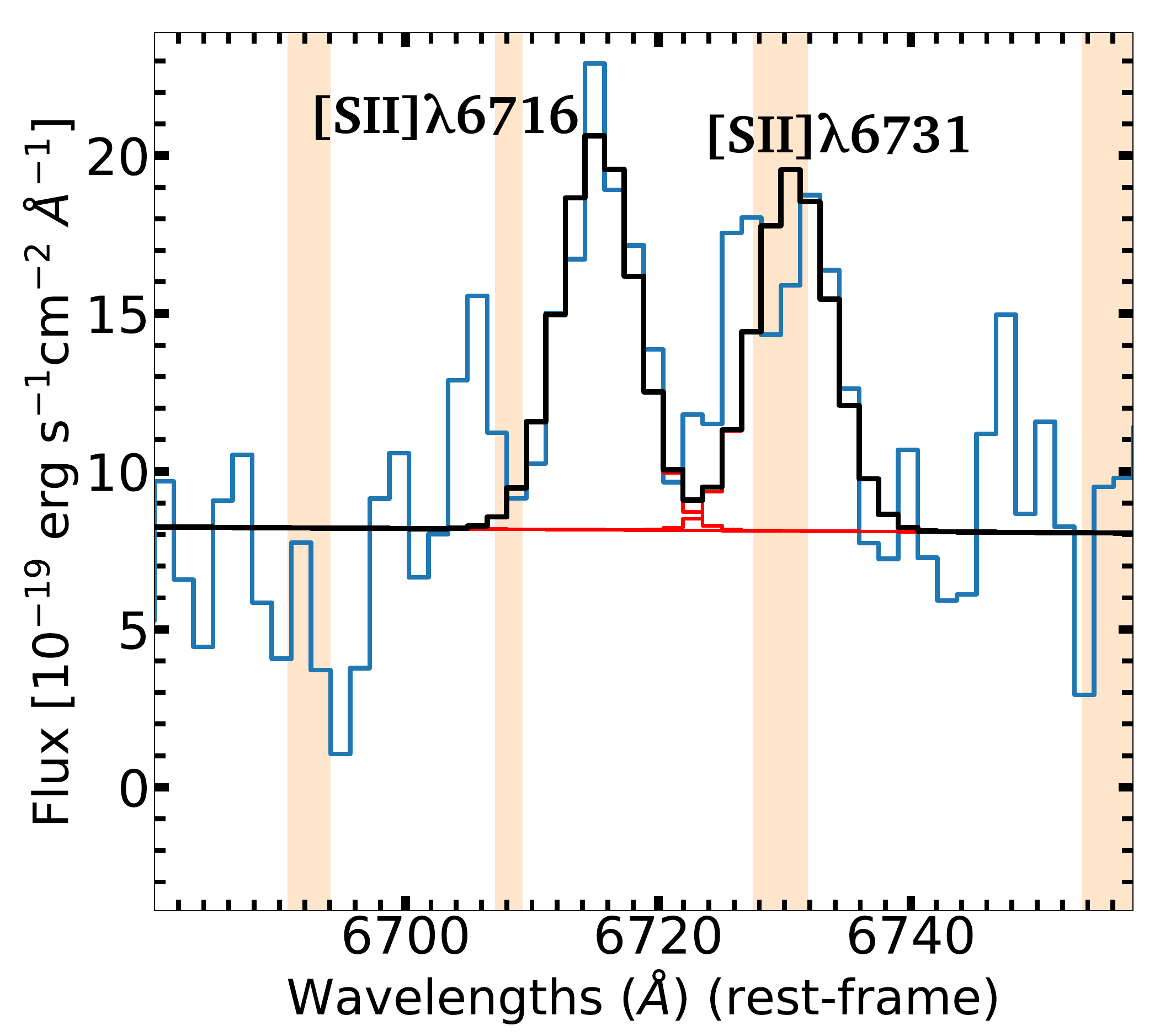}
\end{center}
\caption{Top panel: GNIRS one-dimensional spectrum. Middle panel: line fits of H$\beta$ and [O III] $\lambda \lambda$4959, 5007. Bottom panels: line fits of [NII] $\lambda \lambda$6548,6584 and H$\alpha$ (left) and [S II] $\lambda \lambda$6716,6731 (right, fit (b) of Table~\ref{tab:fit}). The black line represents the model superimposed on the data (blue line), with the single components in red. The vertical bands show the location of the atmospheric lines. The profile of [O III] $\lambda \lambda$4959, 5007 shows a blueshifted wing that was modelled with a broad Gaussian function (FWHM $\sim$ 1270 \kms).}
\label{fig:fit}
\end{figure*}
We used the data of \citet{2008ApJ...677..219K} (see \citealt{2005ApJ...622L..13V} for the details) obtained with the GNIRS spectrograph (\citealt{2006SPIE.6269E..4CE}) at the Gemini South Observatory. 
They consist of a one-dimensional and two-dimensional spectra, publicly available, that cover the range 1.0 - 2.5 $\mu$m with a resolving power $R \sim 1800$.
\begin{table*}
	\caption{Emission line properties derived from the line fitting of the GNIRS spectrum.}
	\centering
	\resizebox*{0.7\textwidth}{!}{
	\begin{tabular}{cccc} 
		\hline
		\hline
       	Line&Flux&Line centroid&FWHM \\
                &($10^{-19}$ erg s$^{-1}$ cm$^{-2}$)&(rest-frame, \AA )&(\kms)\\
        \hline        
        H$\beta$&278 $\pm$ 58 & 4862.4$\pm$0.4 & 264 $\pm$ 63\\
        $\rm{[O III]} \lambda 4959$ (narrow)&936 $\pm$ 125&4958.7$\pm$0.2&394 $\pm$ 40\\ 
       $\rm{[O III]}$ $\lambda$4959 (broad)&501 $\pm$ 157&4949$\pm$2&1170 $\pm$ 389\\ 
       $\rm{[O III]}$ $\lambda$5007 (narrow)&2084 $\pm$ 249&5006.8$\pm$0.2&354 $\pm$ 31\\
       $\rm{[O III]}$ $\lambda$5007 (broad)&1099 $\pm$ 262 &4998$\pm$2&1376 $\pm$ 356\\ 
       \hline       
       $\rm{[N II]}$ $\lambda$6548& 847$\pm$ 107&6547.4$\pm$0.5&756 $\pm$ 78 \\
       H$\alpha$ &2312 $\pm$ 108&6562.6$\pm$ 0.3&604 $\pm$ 30\\
       $\rm{[N II]}$ $\lambda$6584&2120 $\pm$ 149&6583.7$\pm$0.5&756 $\pm$ 78\\  
       \hline       
       $\rm{[S II]}$ $\lambda$6716 (a)&307$\pm$ 58&6715.9$\pm$0.6&319 $\pm$ 79\\
       $\rm{[S II]}$ $\lambda$6731&327 $\pm$ 73&6731.4$\pm$0.6&413 $\pm$ 92\\
       \hline      
       $\rm{[S II]}$ $\lambda$6716 (b)&307 $\pm$ 58&6716.0$\pm$0.5&319 $\pm$ 79\\
       $\rm{[S II]}$ $\lambda$6731&282 $\pm$ 59&6731.5$\pm$0.5&319  $\pm$ 79\\
		\hline
		\hline
		\multicolumn{4}{p{.6\textwidth}}{\textbf{Notes.} 
		Letter (a) refers to the line fit of the \sii\ doublet in which only the centroids ratio was fixed to the theoretical value while letter (b) indicates the fit in which the FWHM of [S II] $\lambda$6731 was matched to the FWHM of [S II] $\lambda$6716 calculated in fit (a).}
	\end{tabular}}
\label{tab:fit}	
\end{table*}
Several emission lines are present such as [O II] $\lambda \lambda$3727, 3729, $\rm{[O III]}$ $\lambda \lambda$4959, 5007, H$\alpha$, H$\beta$, [N II] $\lambda \lambda$6548, 6584 and the [S II] $\lambda \lambda$6716, 6731 doublet (Figure~\ref{fig:fit}, top panel).\\ 
\indent We analyzed the one-dimensional spectrum through line fitting. The analysis was carried out with SPECFIT \citep{1994ASPC...61..437K}, a spectral fitting tool that
runs in the Image Reduction and Analysis Facility (IRAF) environment.
The galaxy emission was modelled with a linear continuum and eight Gaussian components, one for each line. 
We divided the spectrum into two ranges including H$\beta$, \oiii\ (1.5 - 1.7$\mu$m) and H$\alpha$, \nii , \sii\ (2.0 - 2.3 $\mu$m) on which we performed two fits separately.
The ratio between the line centroids of oxygen, nitrogen and sulfur doublets was fixed to the theoretical one (0.990, 0.995 and 0.998 respectively).
We initially used two Gaussian functions to reproduce [O III] $\lambda \lambda$4959, 5007. However, both the lines show a broad, blueshifted wing, detected at 3$\sigma$ and 5$\sigma$ of significance for  [O III] $\lambda$4959 and  [O III] $\lambda$5007, that is not reproduced by one Gaussian. In order to account for these emissions, we performed a second fit adding two other Gaussians (Figure~\ref{fig:fit}, middle panel).
This addition improved significantly the reduced chi-square $\chi^2_{\rm{dof}}$ from 1.23 to 1.01. The new fitted components have a FWHM $\sim$ 1270 \kms , much broader in comparison to those already fitted (FWHM $\sim$ 370 \kms).
The blueshifted component of \5 is also clearly visible in the GNIRS two-dimensional spectrum (broad component in Figure~\ref{fig:pv_iniz}, left panel).\\
\indent In the case of H$\alpha$ and \nii\ we note that the fitted line widths result considerably larger than the FWHM of other elements (Table~\ref{tab:fit}). This could be due either to the overlapping of several sky lines, which may alter the line profiles, or to the presence of blueshifted components, hidden by the line blending, that were not modelled separately (Figure~\ref{fig:fit}, bottom left panel).
The latter interpretation could be supported by the blueshifted emission affecting the H$\alpha$ profile after the subtraction of the \nii\ models from the KMOS spectrum (see section~\ref{sub:kmos} and Figure~\ref{fig:pv_iniz}, right panel).\\
\indent In the \sii\ case, two fits have been performed. If only the centroids ratio is fixed, the FWHM of [S II] $\lambda$6731 results indeed higher than the one relative to [S II] $\lambda$6716 (Table~\ref{tab:fit}, fit (a)). 
However, the [S II] $\lambda$6731 emission is affected by a sky line (Figure~\ref{fig:fit}, bottom right panel) that likely alters its profile. In order to account for this uncertainty, we performed a second fit in which the line widths were matched to the FWHM of [S II] $\lambda$6716 (i.e. the more reliable one) calculated in the first fit.\\
\indent All the parameters derived from line fitting are reported in Table~\ref{tab:fit}. The fitted values are consistent with the ones found by \citet{2005ApJ...622L..13V} despite of the addition of the broad \oiii\ components.\par
We used the fitted quantities to evaluate some global properties of GMASS 0953 like dust obscuration, the SFR and the electron number density.
Assuming a \citet{2000ApJ...533..682C} extinction curve, we estimated the color excess $E(B-V) = 0.8 \pm 0.3$ from the line ratio of H$\alpha$ and H$\beta$. This value indicates that GMASS 0953 is a highly obscured system (see also \citealt{2005ApJ...622L..13V}). Before calculating the color excess, the fluxes of H$\alpha$ and H$\beta$ were corrected for the stellar continuum absorption considering an equivalent width EW = 4 $\pm$ 1 \AA\ \citep{2005ApJ...622L..13V}. Then we derived the SFR$_{\rm{H}\alpha}$ = 498 $\pm$ 33 M$_{\odot}$yr$^{-1}$ using the de-reddened H$\alpha$ through the \citet{1998ARA&A..36..189K} relation, rescaled to a \citet{2003PASP..115..763C} IMF. This value is higher than the one obtained using the infrared luminosity, likely because of the AGN contamination (see also section~\ref{sub:kmos}).\\ 
\indent Finally, we measured the electron number density $n_\text{e}$ from the \sii\ line ratio calculated in the two fits. Both ratios are compatible with an electron number density of 500 cm$^{-3}$ \citep{2006agna.book.....O}, which can vary in the range $100 < n_\text{e}$ [cm$^{-3}] < 2 \times 10^3$ because of the uncertainties in the line ratios. Values of $n_\text{e} \sim 500$\ $\rm{cm}^{-3}$ are typical of AGN narrow line regions. 
However, it is highly reasonable that the galaxy ISM also contributes to it because of the large size of the ionized gas, that extends over the entire galaxy as visible from the SINFONI and KMOS maps (Figure~\ref{fig:maps} and Figure~\ref{fig:HST_disc}).
\subsection{SINFONI integral field spectrum}
\label{sub:sinfoni}
\begin{figure*}
\begin{center}
\includegraphics[scale=0.5]{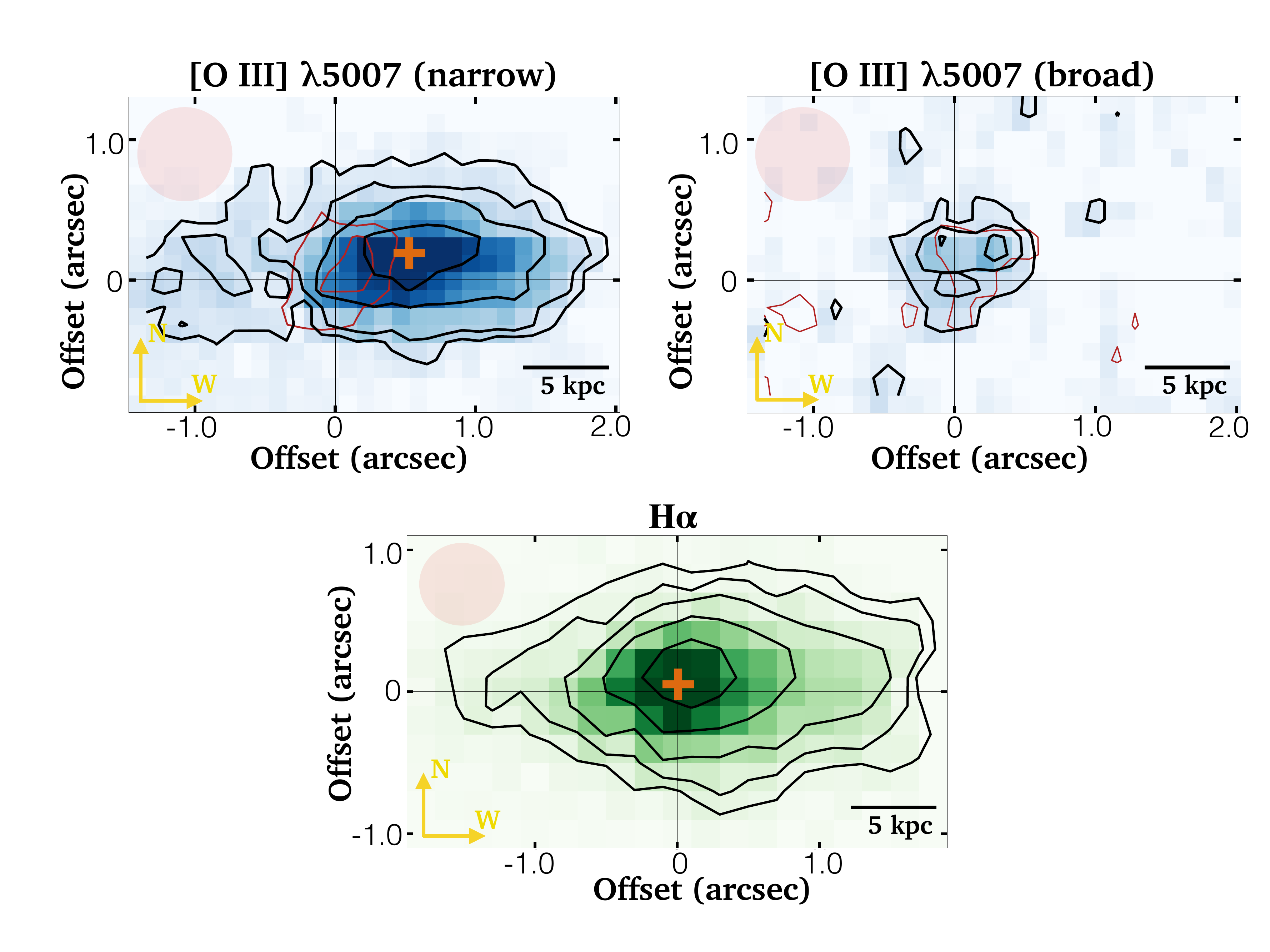}
\end{center}
\caption{Emission line maps extracted from the SINFONI and KMOS data-cubes. Top panels: narrow and broad component of  [O III] $\lambda$5007 (black contours and colors). Flux contour levels are at 3, 6, 12, 24$\sigma$ for the \5 narrow line and at 2, 4, 6$\sigma$ for the \5 broad component.
The red contour shows the continuum emission stacked on the entire spectrum (left panel) and on $N=18$ channels (right panel), i.e. the same number of co-added channels for the \5 blueshifted component. Compared to the center of GMASS 0953 (0$''$ offset in the map) the \5 narrow line morphology appears asymmetric, concentered in the western region of the galaxy, as indicated also by the off-centered emission peak (orange cross); on the other hand the \5 blueshifted line emits close to the nuclear region.
Bottom panel: H$\alpha$ emission. Flux contour levels are at 3, 6, 12, 24, 48$\sigma$. Compared to the \5 emission H$\alpha$ shows a more symmetric morphology, with the emission peak (orange cross) close to the center.
All the maps are centered on the peak of the \textit{HST/H}-band image.
The beam size of SINFONI and KMOS is shown in pink.}
\label{fig:maps}
\end{figure*}
We analyzed the data of GMASS 0953 from the SINS survey (see \citealt{2009ApJ...706.1364F} for the observation details and data reduction) carried out with the SINFONI \citep{2003SPIE.4841.1548E} integral field spectrograph (IFS) at the ESO Very Large Telescope (VLT). They consist in a data-cube covering the \textit{H}-band (1.5 - 1.7 $\mu$m)  with a resolving power $R\sim 2900$ and a channel width of 1.95 \AA\ containing the sky-subtracted data. The visible lines are H$\beta$ and the \oiii\ doublet. The observation is seeing limited with a PSF of 0.7$''$. We verified the SINFONI astrometry in order to properly overlap the emission line maps on the \textit{HST} images of GMASS 0953 as explained in Appendix~\ref{appa}.\\
\indent The SINFONI data-cube (and also the KMOS one) was handled using the software GIPSY  \citep{1992ASPC...25..131V}. We focused in particular on the \5 narrow line fitted in the GNIRS spectrum, which we used for the kinematic modelling due to its high brightness, and on the \5 blueshifted component. 
The total flux map, which shows the \5 narrow line morphology, was derived co-adding $N = 17$ channels after masking out all the pixels with a signal-to-noise ratio $F/\sigma$ < 1 and removing by hand (channel by channel) the instrumental noise peaks present in the field of view. 
We considered as the typical noise in one channel the average standard deviation $\sigma$ =  $1.7 \times 10^{-20}$ erg s$^{-1}$cm$^{-2}$\AA$^{-1}$ of the signal in the channels without emission from the galaxy. The noise of the \5 map was estimated by the relation $\sigma_{\text{TOT}} = \sigma \sqrt{\bar{N}}$, where $\bar{N}$ = 9  is the average number of co-added channels at each pixel position in the map. Due to the masking process, the number of co-added channels is not constant for all the pixels.\\ 
\indent From the map\footnote{We point out that the \5 map was not continuum subtracted because the contributed continuum flux on $N=17$ channels is very low (less of 2$\sigma$ significance).} (Figure~\ref{fig:maps}, top left panel), the \5 narrow line emission appears quite extended (see Figure~\ref{fig:HST_disc} for a visual comparison with the \textit{HST/WFC3} image). Moreover, the flux distribution is very asymmetric, with the western side much brighter than the eastern one. The asymmetry in the flux distrubution is visible also in the GNIRS two-dimensional spectrum (narrow component in Figure~\ref{fig:pv_iniz}, left panel).\\
\indent Then, after a by eye inspection of the whole data-cube, we stacked the channels without emission lines or sky subtraction residuals to obtain the continuum radiated by the galaxy. The continuum emission is produced by the inner region of GMASS 0953 (Figure~\ref{fig:maps}, top left panel) and it originates from the stellar component; however, also a contribution from the obscured AGN is possible. This is indeed less affected by dust extinction at rest-frame optical wavelengths than in the UV, where the nuclear activity of GMASS 0953 does not emerge (see section~\ref{sec:portrait}).
\begin{figure*}
\begin{center}
\includegraphics[scale=0.5]{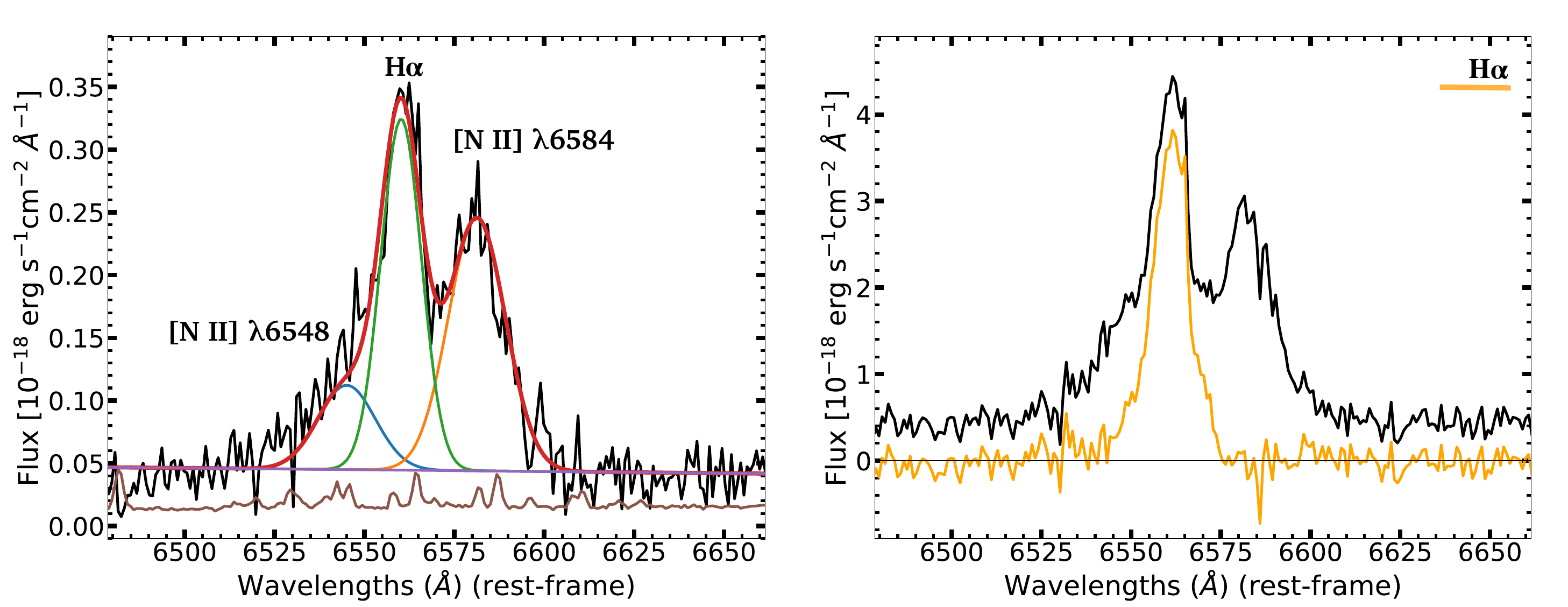}
\end{center}
\caption{H$\alpha$ and \nii\ emission lines (KMOS data). Left panel:  line fit of H$\alpha$ and \nii\ superimposed on the data (black line) in one pixel. The fitted Gaussians are shown in colored lines with the total model in red. The purple line indicates the continuum level while the noise spectrum is shown in brown. Right panel: total H$\alpha$ emission (orange) after the subtraction of \nii\ and stellar continuum. See also Appendix~\ref{appb}.}
\label{fig:kmosFit}
\end{figure*}
In order to properly overlap the SINFONI line maps to the \textit{HST} images of the galaxy (see Appendix~\ref{appa}), 
we evaluated the coordinates of the continuum peak fitting the stacked emission with a two-dimensional elliptical Gaussian.
We observe that there is a shift of $0.58'' \pm 0.24''$ (equivalent to 4.9 $\pm$ 2.0 kpc) between the fitted coordinates and the \5 narrow line peak (cross in Figure~\ref{fig:maps}, top left panel), evaluated also through a two-dimensional Gaussian fit.\\ 
\indent The detection of blueshifted emission with respect to the \5 narrow line, corresponding to the broad component fitted in the GNIRS spectrum (Figure~\ref{fig:fit}, middle panel), was confirmed by the SINFONI data, which also provided its spatial position. Adding $N=18$ channels from 4974 \AA\ to 4995 \AA\ (rest-frame wavelengths), we obtained a 6$\sigma$ detection localized near the galaxy center (Figure~\ref{fig:maps}, top right panel). In order to assess if this emission was truly due to  \5 and not to 
the stellar continuum, which also peaks near the galaxy center, we compared it to the continuum flux of $N=18$ co-added channels,
 \ i.e. the same number of stacked channels for the \5 broad component. These channels were selected at wavelengths close to the oxygen line. As evident from the comparison of the contours, the blueshifted component is significantly brighter than the continuum, hence it traces a genuine emission from the \5 line\footnote{In the [O III] $\lambda$4959 case an emission near the galaxy center was also detected but it was only slightly brighter than the continuum level so we do not report it.}. This component is clearly visible also by looking at the position-velocity (pv) diagram extracted one pixel above the major axis of the galaxy (Figure~\ref{fig:pv_iniz}, middle panel). It appears like an evident off-set emission from the \5 narrow line reaching a velocity difference from the bulk of the galaxy emission of $\sim-1200$ \kms.\\
\subsection{KMOS integral field spectrum}
\label{sub:kmos}
To model the H$\alpha$ kinematics, we employed integral field data from the  KMOS$^{\rm{3D}}$ survey \citep{2015ApJ...799..209W}, which investigates the kinematics of galaxies at $0.7 < z < 2.7$ using the NIR spectrograph KMOS \citep{2013Msngr.151...21S} at the VLT. GMASS 0953 is one of the compact, massive galaxies in the sample of \citet{2018ApJ...855...97W}. 
It was observed with a resolving power $R\sim$ 3965 at H$\alpha$ wavelengths and a spatial resolution of 0.59$''$. The spectrum covers the range 1.95 - 2.3 $\mu$m with a channel width of 2.8 \AA\ and an average noise of 1.6 $\times$ $10^{-20}$ erg s$^{-1}$cm$^{-2}$\AA$^{-1}$ in each channel. 
The visible lines are H$\alpha$, \nii\ and the \sii\ doublet.
For the observation details and the data reduction see \citet{2015ApJ...799..209W}.\\
\indent In order to study the kinematics of H$\alpha$, we carried out a line fit to subtract  the stellar continuum and the nitrogen emission that is partially mixed with the H$\alpha$ line (Figure~\ref{fig:kmosFit}, left panel), as already seen in the GNIRS spectrum (Figure~\ref{fig:fit}, bottom left panel).
For details about the line fit, that we performed pixel by pixel, see Appendix~\ref{appb}.
After subtracting the \nii\ models from the original spectrum, a new data-cube containing only the H$\alpha$ emission was built (Figure~\ref{fig:kmosFit}, right panel). We then stacked the channels with the line ($N = 25$) to obtain the total flux map (Figure~\ref{fig:maps}, bottom panel).
In comparison to the \5 emission, the H$\alpha$ morphology appears much less asymmetric, with the line peak closer to the galaxy center.
The central position of the peak may support the substantial contribution to the line from the AGN, which explains the discrepancy between the H$\alpha$ derived SFR and the one inferred from the IR luminosity (see section~\ref{sub:gnirs}).
\section{Kinematic modelling}
\label{sec:kine}
The SINFONI and KMOS data were used to perform a study of the kinematics of the \5 narrow line and H$\alpha$.
The two lines show a large-scale velocity gradient that we analyzed in details (Figure~\ref{fig:pv_iniz}).
In particular, we worked under the hypotesis that the observed velocity gradients are due to gas rotation and we modelled the line emission as if it were produced by a rotating disc.
In section~\ref{sub:massiveOf} we also discuss an alternative interpretation of the velocity gradient.
\subsection{Signatures of an ionized disc}
\begin{figure*}
\begin{center}
\includegraphics[scale=0.52]{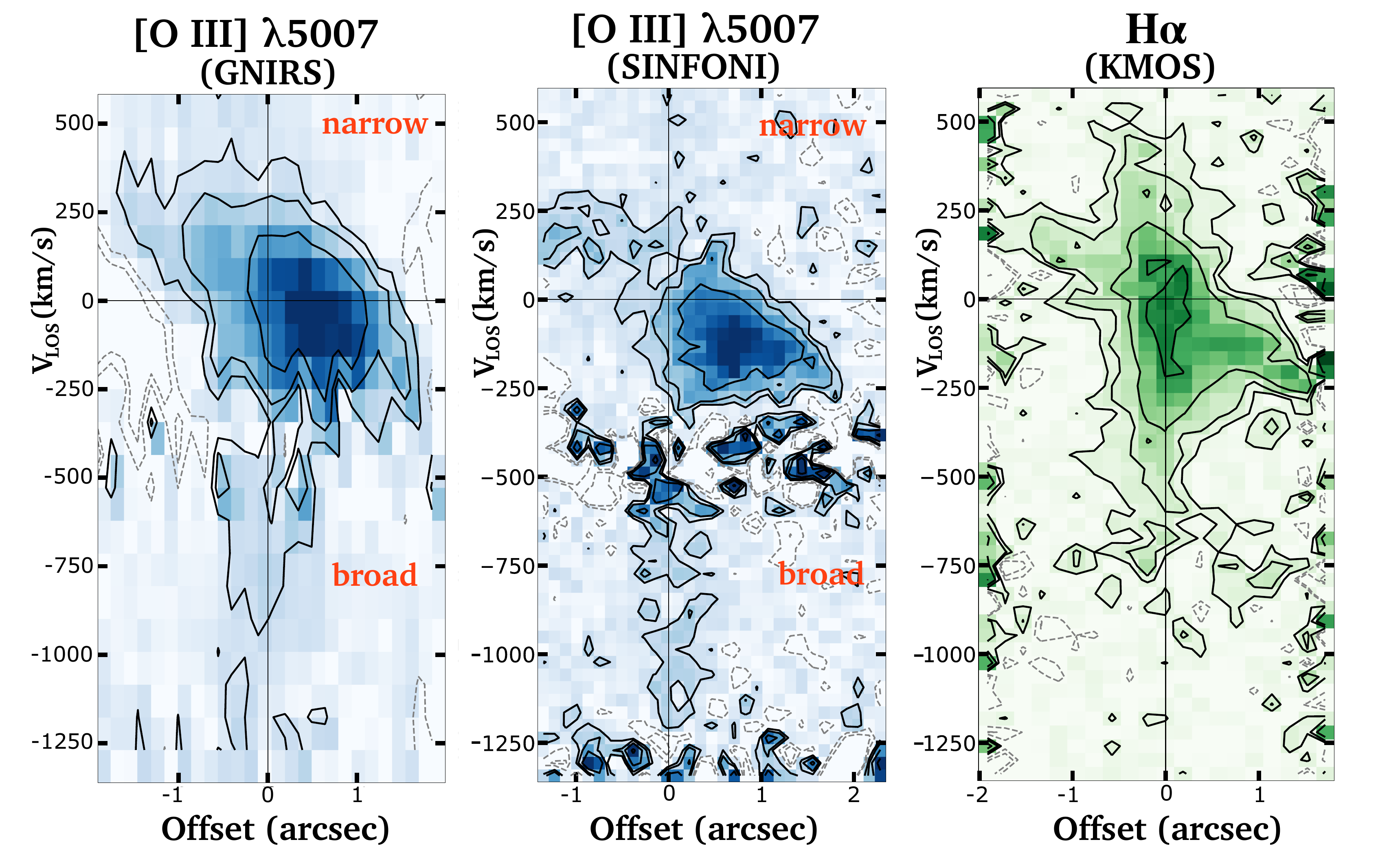}
\end{center}
\caption{Position-velocity (pv) diagrams of \5 and H$\alpha$ roughly along the major axis of GMASS 0953 extracted from the GNIRS two-dimensional spectrum (left), SINFONI (middle) and KMOS (right) data-cubes. Flux contour levels are at $\pm$2, $\pm$4, 8, 16$\sigma$. Both the \5 and H$\alpha$ lines present a large-scale velocity gradient that extends up to 500\kms at the external radii and goes far beyond in the central region. However, in the H$\alpha$ case, a contamination from the \nii\ residuals is possible. The central excess up to $\sim -1200$ \kms\ presented by \5 in both the GNIRS and SINFONI pv diagrams corresponds to the broad blueshifted component fitted in the GNIRS one-dimensional spectrum.}
\label{fig:pv_iniz}
\end{figure*}
As suggested by an inspection of the GNIRS two-dimensional spectrum (Figure~\ref{fig:pv_iniz}, left panel), the \5 narrow line emission appears tilted along the slit, showing a large-scale velocity gradient that could be due to gas rotation. 
The emission extends over $\sim$ 3$''$, corresponding to 25 kpc.
This large-scale gradient has been confirmed by the pv diagram of \5 extracted roughly along the major axis of GMASS 0953 (see Figure~\ref{fig:hst_h}, top panel) using the SINFONI data (middle panel in Figure~\ref{fig:pv_iniz}). 
Before the extraction, the data-cube channel separation was converted into velocity after determining $\lambda_{\rm{rest}} = (\lambda^{B}_{50} + \lambda^{R}_{50}) / 2$, i.e. the average value between the wavelengths $\lambda_{50}$ where the flux of the global \5 profile is 50\% of the peak, with 'B' and 'R' that refer to the blue and red side of the line profile respectively.
This quantity was set to zero for simplicity and allowed us to fix the channel velocity spacing to $\Delta v$ = 36.6 \kms\ at the \5 wavelengths\footnote{For H$\alpha$ the channel velocity spacing is 39.8 \kms. The $\lambda_{\rm{rest}}$ value was used also to convert the wavelengths of the GNIRS two-dimensional spectrum into velocity.}.\\
\indent The emission of the \5 narrow line covers about 500 \kms with the western, bright side of the galaxy moving in our direction while the eastern, dimmer part is globally redshifted. 
Because of an atmospheric line, it is not clear if the galaxy emission reaches velocity higher than $-300$ \kms, which would be hidden by the sky-subtraction.
The large-scale gradient extracted from the SINFONI data is consistent with the observed velocity in the GNIRS two-dimensional spectrum, even if the latter shows a slightly higher velocity for the receding half probably because of the better sensitivity and lower spectral resolution.\\ 
\indent As we noted in section~\ref{sub:sinfoni}, in the central region of GMASS 0953 the \5 gradient extends up to $\sim -1200$ \kms\ (broad component in Figure~\ref{fig:pv_iniz}, middle panel). The interpretation of this blueshifted emission, which is not due to gas rotation, will be discussed in section~\ref{sec:feedback}.\\
\indent Interestingly, a large-scale velocity gradient is also shown by H$\alpha$ (Figure~\ref{fig:pv_iniz}, right panel). The line shows a slightly asymmetric morphology, not marked as in the \5 case, with the redshifted half more clearly visible. The emission from the external regions shows the same velocity gradient of the \5 line for both sides while, in the inner part, it extends up to 500 \kms\ for the red half and goes far beyond on the blue side. This blueshifted emission, which extends on a smaller scale ($\sim 0.8''$, corresponding to $\sim 6.7$ kpc) with a velocity up to $-700$ \kms, highlights the presence of a possible blueshifted component also in H$\alpha$, though we point out a probable contamination from the residuals of the [N II] $\lambda$6548 subtraction.
Evidence of a blueshifted component necessary to reproduce H$\alpha$ and the \nii\ doublet was found by \citet{2014ApJ...787...38F} and  \citet{2014ApJ...796....7G}, who interpreted it as a nuclear outflow linked to the AGN activity of the galaxy.\\
\indent A velocity gradient, extending in the same direction of the \5\ and H$\alpha$ large-scale observed gradients, is also shown by the CO(J=6-5) line, which traces a high density, compact molecular disc hosted by GMASS 0953 (\citealt{2018MNRAS.476.3956T}). This fact suggests that the galaxy harbors a multi-phase disc that includes both an ionized and a molecular component (see section~\ref{sub:multiDisc}).
\subsection{Three-dimensional disc modelling}
\label{sub:barolo}
\begin{figure}
\begin{center}
\includegraphics[scale=0.49]{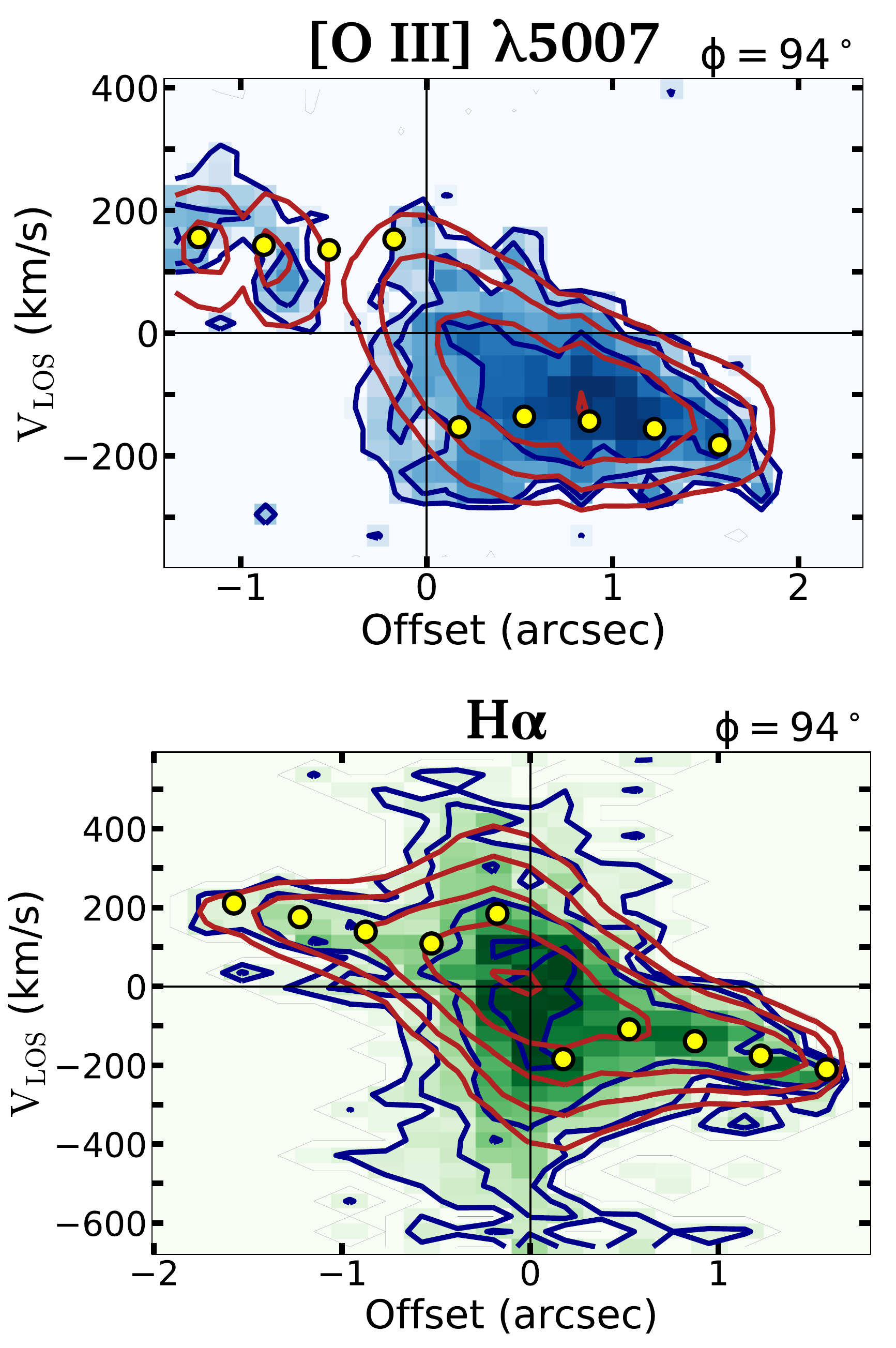}
\end{center}
\caption{Position-velocity diagrams of the \5 narrow line (top panel) and H$\alpha$ (bottom panel) extracted along the major axis ($\phi = 94$\degr) of GMASS 0953.
Blue contour and colors refer to the data while the $^{\rm{3D}}$\textsc{Barolo} disc model is marked in red.
Flux contour levels are at  $\pm$2, 4, 8, 16, 32$\sigma$. The yellow dots show the rotation curve projected along the line of sight.}
\label{fig:pv_norm}
\end{figure}

We interpreted both the \5 and H$\alpha$ large-scale velocity gradient as due to a rotating disc that we modelled with the publicly available software $^{\rm{3D}}$\textsc{Barolo} \citep{2015MNRAS.451.3021D}. This is an algorithm based on the tilted-ring model \citep{1974ApJ...193..309R} that derives rotation curves through a three-dimensional modelling of data-cubes.
In this approach, the rotation velocity and the velocity dispersion are computed directly comparing the observed data-cube with an artificial one that simulates an IFS observation, avoiding the extraction of the two-dimensional velocity field.
The latter is indeed affected by beam smearing \citep{1987PhDT.......199B}, a problem due to the low spatial resolution (e.g. atmospheric seeing) that spreads the line emission within a region on the adjacent ones. The result is that the observed velocity gradients are artificially flattened with a consequent increase of the line broadening, which makes the kinematical parameters derived from the two-dimensional velocity field less reliable
(see \citealt{2015MNRAS.451.3021D} and \citealt{2016A&A...594A..77D} for more details).
On the other hand, a three-dimensional modelling is not affected by beam smearing because the PSF of the observation is taken into account through a convolution step, performed to build mock data with the same spatial resolution of the observed ones.\\ 
\indent To reproduce the data, the code builds a three dimensional disc model, formed by $N$ rings of width $W$, for which the parameters that minimize the residuals between the data-cube and the model-cube are calculated. These parameters, both geometrical and kinematical, are the rotation velocity $V(R)$, the velocity dispersion $\sigma (R)$, the inclination angle $i$ (90\degr\ for edge-on discs), the position angle $\phi$ of the major axis, the dynamical center $(x_0, y_0)$, the systemic velocity $V_{\text{sys}}$, the disc scale height $z_0$ and the gas surface density $\Sigma$.\\
\indent We applied the algorithm to \5 and H$\alpha$ independently, fixing all the rings parameters except $V(R)$ and $\sigma(R)$. The galaxy was divided into five rings with radii from 0.175$''$ (1.5 kpc) to 1.575$''$ (13.2 kpc) and a width of 0.35$''$, corresponding to about half the PSF. 
This choice entails that $V(R)$ and $\sigma (R)$ of a ring are not fully independent of the fitted values for the closer ones. The inclination angle $i$ was fixed to 75\degr for both the lines (see Appendix~\ref{appc}) while we chose $\phi = 94$\degr\ after producing pv diagrams for different position angles and considering the one with the most extended emission.
For the dynamical center we used the coordinates of the emission peak in the \textit{HST/H}-band image while we left the disc scale height $z_0$ to the default value (150 pc): due to the low spatial resolution the disc scale height is unresolved, hence the thickness has a negligible effect.
Finally, we adopted for the gas surface density $\Sigma$ a locally averaged normalization, in which the flux in each pixel of the disc model is equalled to the flux in the corresponding pixel in the data.
In this way the inhomogenities that may affect a ring, as in the \5 case, can be reproduced.\\
\indent After setting the disc parameters as described above, we launched $^{\rm{3D}}$\textsc{Barolo} obtaining two best-fit model-cubes that simulate the \5 and H$\alpha$ line emission. 
In order to compare the disc model with the data, we inspected the pv diagrams extracted along the major axis (Figure~\ref{fig:pv_norm}).
The observed \5 and H$\alpha$ emission is well reproduced by the model. 
However, some differences are present. For example, the model slightly overestimates the line broadening for both \5 and H$\alpha$ in the outer rings ($R\sim 1.8''$) of the approaching side.
The disc model also does not reproduce the emission in the central region, with possible repercussions on $V(R)$ and $\sigma (R)$ of the internal rings. This effect is particularly prominent in the H$\alpha$ case.\\
\begin{figure}
\begin{center}
\includegraphics[scale=0.48]{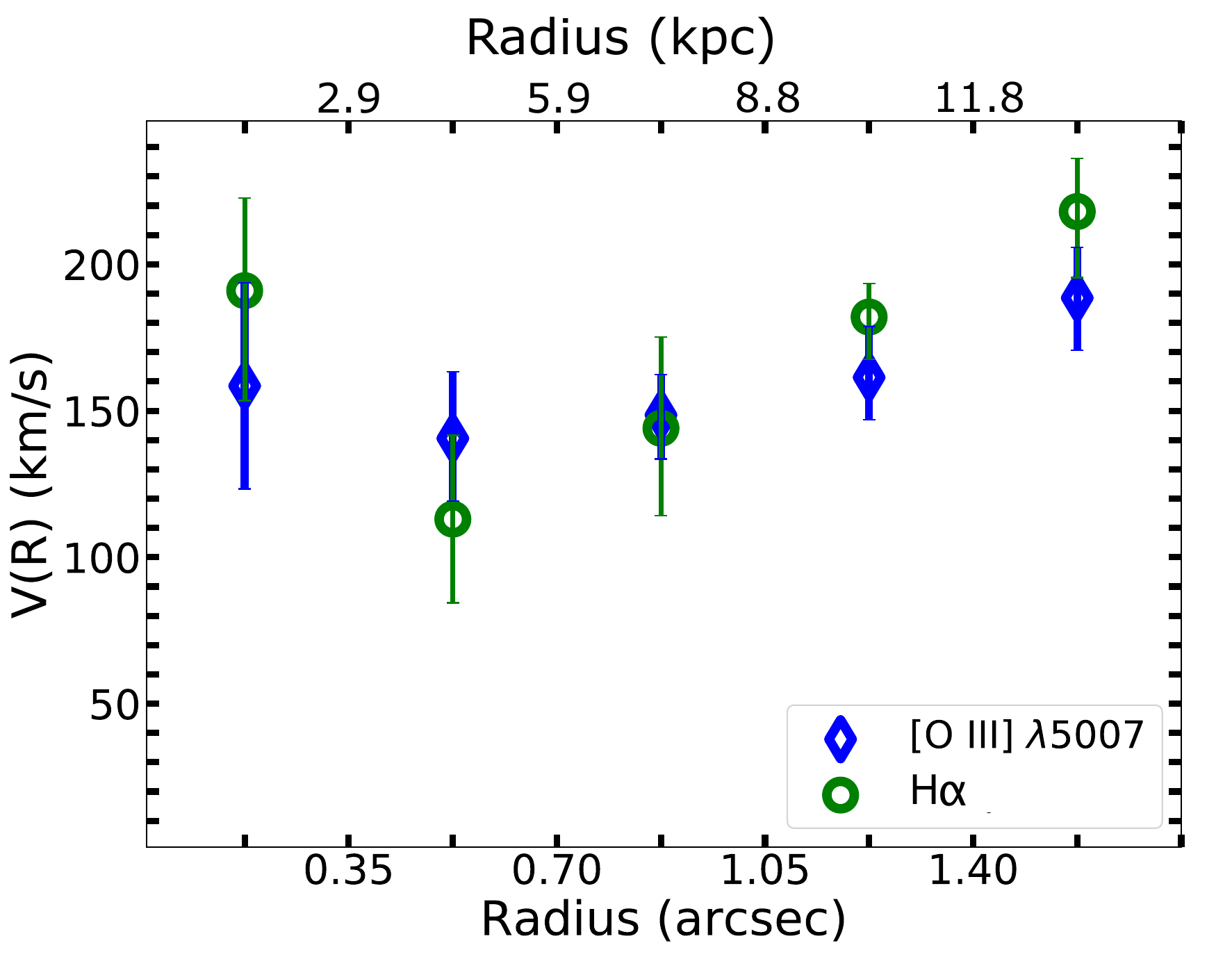}
\end{center}
\caption{Rotation curve of \5 (blue diamonds) and H$\alpha$ (green circles). The rotation velocity decreases between the first two rings and increases going outwards. The $V(R)$ values are in agreement within the errors for both the lines in all the rings.}
\label{fig:rotcurve}
\end{figure}
\indent The fitted rotation curves nicely show consistent values of $V(R)$ for the two lines in each ring (Figure~\ref{fig:rotcurve}).
The rotation velocity decreases between the first two rings and appears to slightly increase going outwards for both \5 and H$\alpha$.\\
\indent We assessed the reliability of the best-fit velocities simulating new model-cubes in which we varied the fitted $V(R)$ in order to visually verify if different values of the rotation velocity worsened or improved the pv diagram reproduction (see \citealt{2009A&A...493..871S} for the details about a similar "trial-and-error" procedure in this context). 
This test was carried out adopting an azimuthal surface density profile for $\Sigma$ instead of the local normalization because in this way the simulated pv diagrams present a smoothed and symmetric emission that makes it more straightforward to see variations in their shape.
For both \5 and H$\alpha$ we found that the shape of the pv diagram is not sensitive to large variations of $V(R)$ for the first two rings, meaning that the associated errorbars are underestimated. On the other hand, we consider reliable the fitted rotation velocity and errors of the external three rings as significantly lower or higher values of $V(R)$ would not reproduce the data.\\
\indent We repeated the test for the velocity dispersion $\sigma (R)$. The evaluation of this parameter is particularly complex because of the low spatial and spectral resolution of our data and the relatively high inclination of the galaxy. We could derive only an upper limit of $\sigma (R) < 160$ km s$^{-1}$ for \5 and  $\sigma (R) < 90$\kms\ for H$\alpha$ at large radii, which we obtained rejecting the $\sigma (R)$ values that do not reproduce the pv diagram.\\
\indent After this analysis, as a representative rotation velocity of the disc, we took the weighted average $V(R)$ of \5 and H$\alpha$ of the last ring ($R\sim 13$ kpc), which amounts to $V_{\rm{ion}} = 203^{+17}_{-20}$ \kms. The associated errors were estimated considering both the fitted ones and the uncertainties due to the inclination angle (see Appendix~\ref{appc}).
\subsection{An alternative interpretation of the velocity gradient}
\label{sub:massiveOf}
\begin{figure*}
\begin{center}
\includegraphics[scale=0.39]{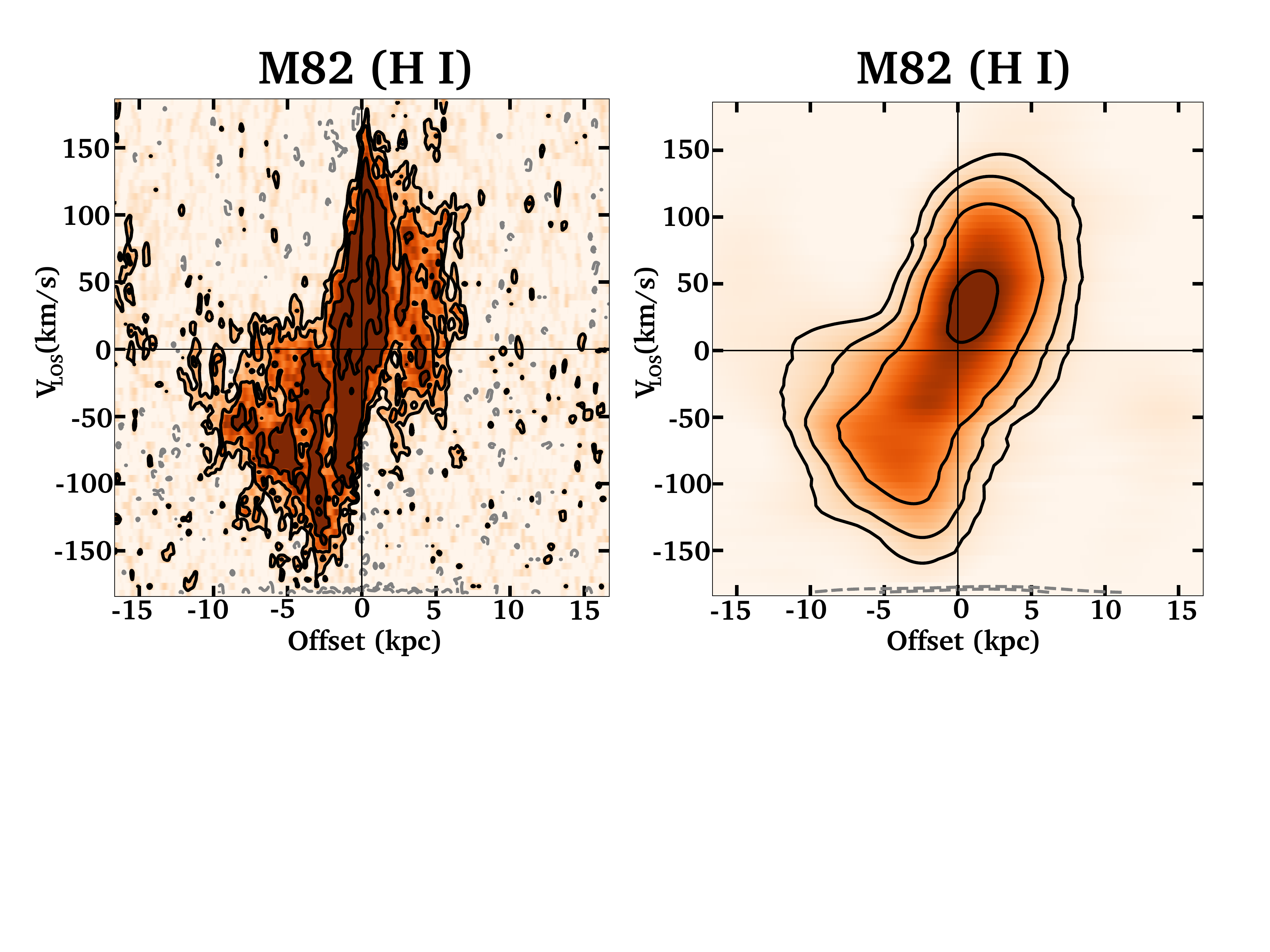}
\end{center}
\caption{Position-velocity diagrams along the HI wind in the local starburst M82 (Very Large Array and Green Bank Telescope data; \citealt{2018ApJ...856...61M}). Left panel: original data with a spatial resolution of 0.4 kpc. Right panel: smoothed data with a spatial resolution of 4.65 kpc. The smoothed emission appears roughly similar to the observed velocity gradient of H$\alpha$ and \5 in GMASS 0953 suggesting that the interpretation of gas velocity gradients may be ambiguous when using data of low spatial resolution.}
\label{fig:m82}
\end{figure*}
In section~\ref{sub:barolo} we modelled the kinematics of \5 and H$\alpha$ under the hypotesis that the velocity gradient of the two lines was produced by a rotating disc.
This is the more straightforward assumption we can do given the limited spatial resolution of our data.
However, velocity gradients in the gas kinematics can also be originated by other phoenomena. In particular, galactic-scale winds can produce velocity gradients similar to that observed in GMASS 0953, with the emission coming from the approaching/receding cone blueshifted and redshifted respectively.
This can be seen in observations of local galaxies.
One of the best studied cases of galactic winds in the nearby Universe concerns the starburst galaxy M82. The wind, due to vigorous star formation in the centre, has been observed in several gas phases (e.g. \citealt{1989ApJ...336..722S, 1998ApJ...493..129S, 2015ApJ...814...83L}).
Recently, \citet{2018ApJ...856...61M} investigated the kinematics of the M82 wind using HI observations resolved on the scale of 0.4 kpc. The wind extends over 10 kpc from the galaxy center and the observed pv diagram (Figure~\ref{fig:m82}, left panel) presents some interesting analogies with those of \5 and H$\alpha$ in GMASS 0953 (Figure~\ref{fig:pv_iniz}). Clearly, a direct comparison of the pv diagrams is difficult because of the different spatial resolution of the two observations. Therefore, we smoothed the M82 data in order to have the same number of resolution elements of GMASS 0953 along the pv diagram. This corresponds to a new spatial resolution of 4.65 kpc. We found that the analogy persists after the smoothing (Figure~\ref{fig:m82}, right panel) in terms of
the global shape of the pv diagram, the asymmetric flux distribution and the observed declining velocity from the inner to the outer regions.  
Despite these similarities, the above comparison is not trivial at least for two reasons: i) we are comparing two different gas phases, i.e. ionized vs neutral, which have different physical properties; ii) we are comparing a local starbust galaxy without AGN activity with a high-redshift active galaxy.  
Keeping this in mind, our experiment suggests that when looking at the pv diagram of \5 and H$\alpha$ a wind interpretation cannot be ruled out with our data.
However, the wind scenario appears quite unlikely in GMASS 0953 since the wind would move along the projected major axis of the galaxy and not in a perpendicular direction, as observed in other low and high-redshift sources (\citealt{1990ApJS...74..833H, 2018MNRAS.tmp.1721L}). Moreover, the connection between a putative large-scale wind and the outflow described in section~\ref{sec:feedback} would remain unclear.
\section{The multi-phase disc of GMASS 0953}
\label{sub:multiDisc}
\begin{figure*}
\begin{center}
\includegraphics[scale=0.7]{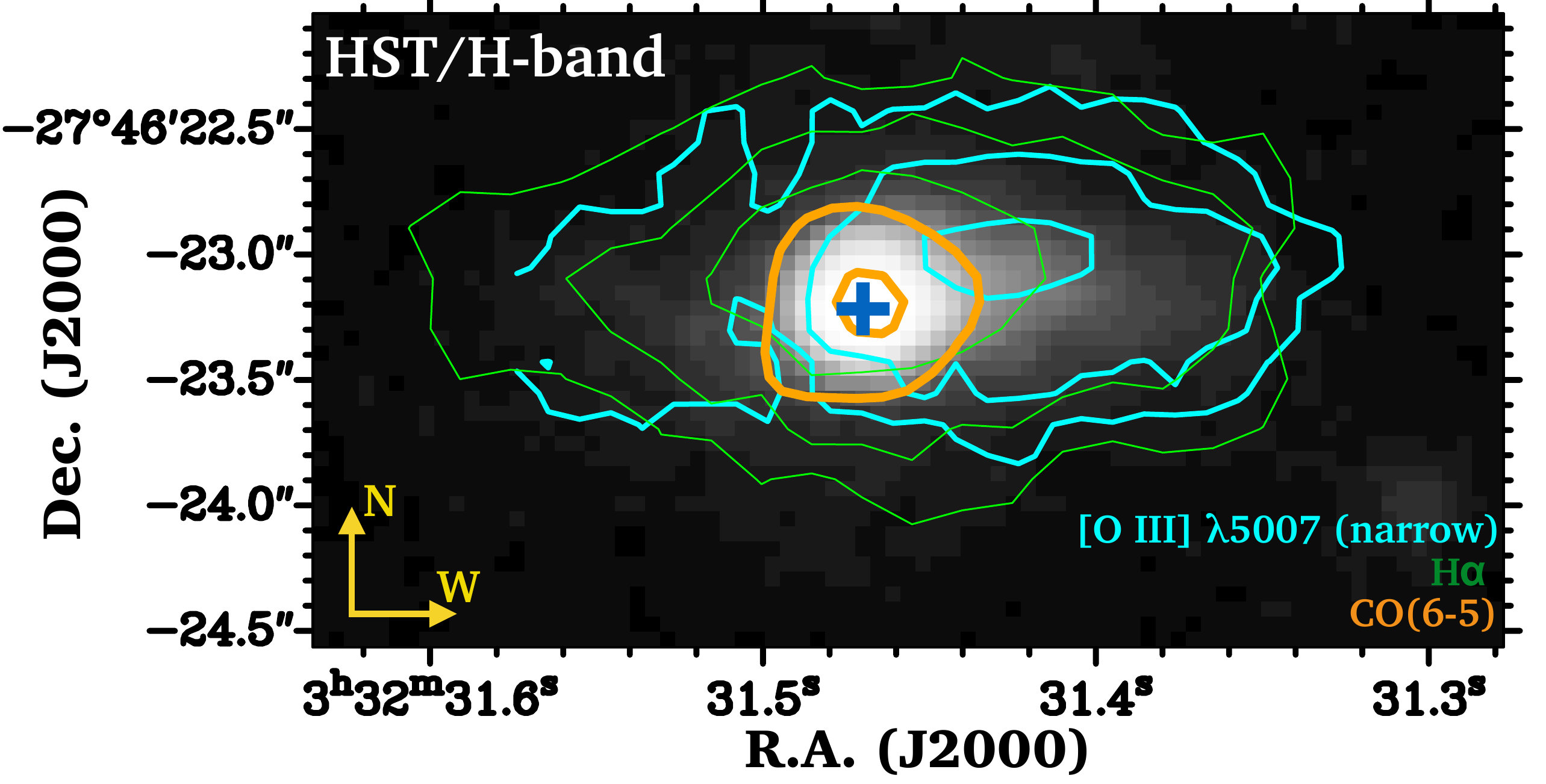}
\end{center}
\caption{The multi-phase gas emission of GMASS 0953 overimposed on the \textit{HST}/WFC3 (\textit{H}-band) image. The contours refer to the \5 narrow line (cyan), H$\alpha$ (green) and CO(J=6-5) (orange, ALMA data; \citealt{2018MNRAS.476.3956T}). The lowest contour level is at 3$\sigma$. The cross indicates the galaxy center.}
\label{fig:HST_disc}
\end{figure*}

From the kinematics modelling presented in section~\ref{sub:barolo}, it emerges that GMASS 0953 may host an ionized disc extending up to $R \sim 13$ kpc with a rotation velocity $V_{\rm{ion}} = 203^{+17}_{-20}$ \kms\ at the largest radius. The existence of the disc is highlighted by two lines analyzed independently, which show two consistent rotation curves.
The H$\alpha$ kinematics of GMASS 0953 was modelled also by \citet{2018ApJ...855...97W}, who found a rotation curve that is broadly consistent with our estimate at large radii, though a direct comparison between the two rotation curves is made difficult by the differences in the adopted disc modelling and beam smearing correction.\\
\indent Rotation does not concern only the ionized phase of the ISM. According to \citet{2018MNRAS.476.3956T}, GMASS 0953 harbors a very compact ($R_{\rm{CO}} = 0.75 \pm 0.25$ kpc) molecular disc traced by the CO(J=6-5) line, with a rotation velocity $V_{\rm{CO}}=$ 320$^{+ 92}_{-53}$ \kms , i.e. higher although compatible at 2$\sigma$ with our H$\alpha$ inner value.
All the three lines show velocity gradients with consistent position angle that are aligned with the major axis of the stellar component. The latter represents possibly a nearly edge-on disc that is visible in the \textit{HST} image of the galaxy (Figure~\ref{fig:HST_disc}). 
These findings suggest that GMASS 0953 hosts a multi-phase disc, with rotation affecting the galaxy from small to large scales and peaking in its central regions.
The joint existence of a cold and an ionized disc in high-$z$ massive galaxies was also found by \citet{2018ApJ...854L..24U} and \citet{2019ApJ...871...37H}.
In the very inner region of GMASS 0953, we considered $V_{\rm{CO}}=$ 320$^{+ 92}_{-53}$ \kms\ of the molecular disc a reliable measure of the rotation velocity. As we discussed in section~\ref{sub:barolo}, the internal points of the \5 and H$\alpha$ rotation curve are dominated by large uncertainties and they could be compatible with the CO(J=6-5) rotation.
Declining rotation curves in the central part of galaxies are observed both in local and distant objects \citep{2006AJ....132.1426S, 2007MNRAS.376.1513N, 2008AJ....136.2648D, 2017Natur.543..397G}. In particular, in local galaxies they have been clearly associated to the presence of large bulges (e.g. \citealt{2016AJ....152..157L}).\\
\indent The ionized gas also provides the $V(R)$ value on a larger scale, hence we can use it to estimate a lower limit to the dynamical mass of the galaxy. Under the hypothesis of pure circular motions, the dynamical mass contained within the radius $R \sim 13$ kpc amounts to $M_{\rm{dyn}} = RV_c^2/G > (1.3^{+ 0.2}_{- 0.3}) \times 10^{11}\ \rm{M}_{\odot}$ where, as the circular velocity $V_c$, we considered the $V_{\rm{ion}}$ value. Our estimate constitutes only a lower limit to the dynamical mass because the circular velocity $V_c$ may be larger than the fitted rotation velocity $V_{\rm{ion}}$. To estimate $V_c$ both H$\alpha$ and \5 may require an asymmetric drift correction \citep{2008gady.book.....B} that is probably negligible for CO(J=6-5) because of the gas lower temperature and velocity dispersion. We have not attempted to apply this correction due to the large uncertainties of our velocity dispersion (see \citealt{2017MNRAS.466.4159I} for the details). The lower limit for the dynamical mass results consistent with the stellar mass evaluated using SED fitting (see section~\ref{sec:portrait}) and also with the upper limit of \citet{2009ApJ...706.1364F}.\\
\indent It is interesting to see where GMASS 0953 is located on the baryonic Tully-Fisher relation \citep{2000ApJ...533L..99M}, which connects the baryonic content $M_{\rm{b}}$ to the circular velocity. 
\begin{figure*}
\begin{center}
\includegraphics[scale=0.7]{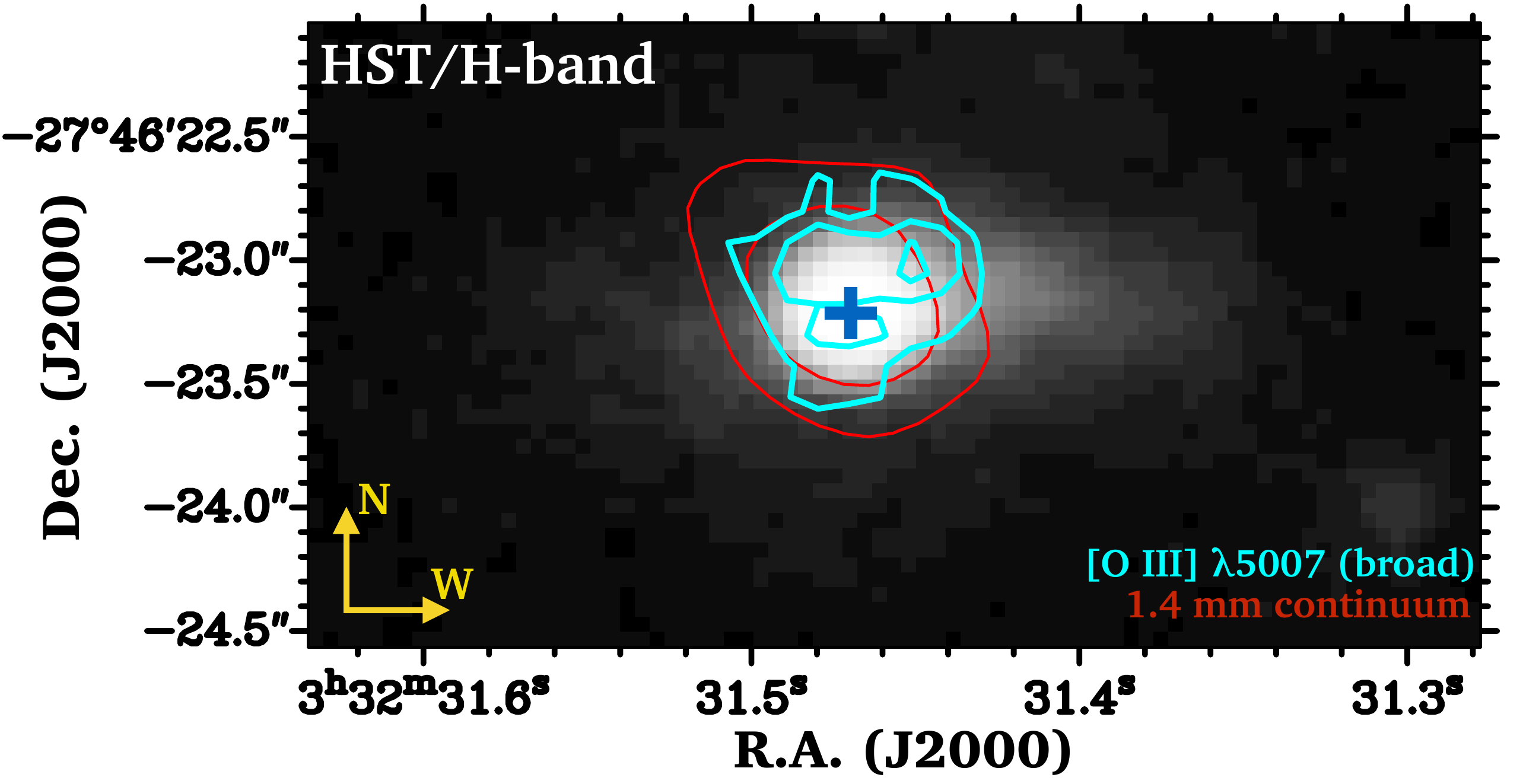}
\end{center}
\caption{Comparison between the \5 blueshifted line (cyan) and the dust continuum (red, ALMA data; \citealt{2018MNRAS.476.3956T}). The two emissions are both concentrated near the center of GMASS 0953, marked with a cross.}
\label{fig:alma}
\end{figure*}
The baryonic content $M_{\rm{b}}$ of the galaxy can be evaluated adding the stellar mass $M_{\star}$ to the gas mass $M_{\rm{g}}$. To compute the gas budget we considered the molecular gas content $M_{\rm{H}_2}$ evaluated by \citet{2018MNRAS.476.3956T} and the ionized gas mass $M_{\rm{ion}}$ derived from the \5\ narrow line luminosity (see Appendix~\ref{appd}). However, the latter amounts to $M_{\rm{ion}} \simeq 8.5 \times 10^{6}\ \rm{M}_{\odot}$ if we use the observed \5 flux or to  $M_{\rm{ion}} \simeq 2.2 \times 10^{8}\ \rm{M}_{\odot}$ when correcting for dust extinction, which is in both the cases negligible respect to the molecular gas mass $M_{\rm{H}_2} = (3.24 \pm 0.60)\times 10^{10}\ \rm{M}_{\odot}$ \citep{2018MNRAS.476.3956T}. We hence computed the baryonic mass as $M_{\rm{b}} \simeq M_{\star} + M_{\rm{H}_2} = (1.47 \pm 0.12) \times 10^{11}\ \rm{M}_{\odot}$. For the circular velocity we used again $V_{\rm{ion}}$ at 13 kpc.
If we put these two values on the local baryonic Tully-Fisher relation of \citet{2016ApJ...816L..14L}, we find that GMASS 0953 is well included into the 0.11 dex scatter of the $z=0$ relation.
This result suggests that massive galaxies that are already relaxed at $z \gtrsim 2$ may exist, in agreement with the findings of \citet{2018MNRAS.tmp.1721L}. 
\section{Probing the AGN feedback in action}
\label{sec:feedback}
\subsection{Evidence of an AGN-driven ionized outflow}
\label{sub:ionized}
We know, from what we discussed in section~\ref{sec:data}, that the \5 line consists of a narrow component, which we used to model the kinematics, and a broad blueshifted wing fitted in the GNIRS spectrum that is not ascribable to the rotating disc (Figure~\ref{fig:fit} and Figure~\ref{fig:pv_iniz}). The emitting region appears concentrated near the nucleus of GMASS 0953 (Figure~\ref{fig:maps}, right panel).\\ 
\indent It is interesting to evaluate the size and the velocity of the \5 blueshifted emission.
We computed the size of the emitting region through a two-dimensional Gaussian fit on the SINFONI map.
The emission appears barely resolved, with a FWHM of the major axis of  $\sim 0.8''$, corresponding to 6.7 kpc.
The velocity can be evaluated from the line fit of the GNIRS spectrum. We measured the velocity shift between the line centroids of the broad and the narrow \5 line to be $\Delta v = - 535 \pm 152$ \kms.\\
\indent Line velocity offsets in galaxies are often connected to outflowing gas due to stellar winds or supernova explosions. However, the typical offset in these cases amounts to $\vert\Delta v\vert \sim 100$ \kms  (e.g. \citealt{2003ApJ...588...65S, 2010ApJ...717..289S, 2012A&A...539A..61T}), which is considerably lower than our value. Conversely, velocity offsets $\Delta v \sim -500$ \kms\ have been found in \5 blueshifted emission extending on several kpc in many high-$z$ active galaxies (e.g. \citealt{2012A&A...537L...8C, 2012MNRAS.426.1073H, 2015A&A...580A.102C, 2015MNRAS.446.2394B}).
These high-velocity kpc-scale emissions are actually interpreted as robust tracers of AGN-driven outflows and represent an observational evidence of the AGN feedback in action.
They are also characterized by line widths $> 1000$ \kms, compatible with the FWHM $\sim$ 1400 \kms\ of the \5 blueshifted line detected in our data.
Finally, also the emitting region, located close to the nucleus GMASS 0953, yet supports the AGN activity as the main engine of gas expulsion in this galaxy.\\ 
\indent The \5 blueshifted line therefore highlights the presence of an AGN-driven ionized outflow that affects GMASS 0953, suggesting that the AGN feedback is likely at work in this galaxy.
A further support to the presence of outflowing material in the ionized phase may come from the blueshifted H$\alpha$ emission in the central part of the galaxy visible in the pv diagram (Figure~\ref{fig:pv_iniz}, right panel; see also \citealt{2014ApJ...787...38F, 2014ApJ...796....7G}).
Moreover, the  FORS2 rest-frame UV spectrum of GMASS 0953 presents absorption lines (e.g. C II $\lambda$1334, Si $\lambda$1260) that are blueshifted of $- 800$ \kms\ respect to the systemic velocity \citep{2013ApJ...779L..13C}. This fact suggests the presence of outflowing material also in a lower ionization state than that traced by \5 and H$\alpha$. Finally, there is a tentative evidence of a blueshifted wing affecting also the CO(J=6-5) line with $\Delta v\sim -700$ \kms\ with respect to the main component, which might be due to outflowing material in the molecular phase \citep{2018MNRAS.476.3956T}.
However, the low significance of the detection (2.5 $\sigma$) prevents us to make a reliable analysis of this component hence, to state the effects of gas expulsion on the galaxy, we considered the ionized phase only.
\subsection{Impact of the outflow on the host galaxy}
\subsubsection{Outflow rate}
Because of the presence of outflowing gas, it is interesting to evaluate the mass of the expelled material per unit time (i.e. the outflow rate) in order to compare it with the SFR of GMASS 0953. 
Many studies observed AGN-driven winds in local galaxies with outflow rates that may exceed the SFR \citep{2011ApJ...729L..27R, 2014A&A...562A..21C} and influence the star formation activity by removing large amounts of gas.\\
\indent We followed the approach of \citet{2012A&A...537L...8C} and \citet{2015A&A...580A.102C}, who derived the outflow rate of the ionized gas from the luminosity of the \5\ blueshifted emission (all the calculations are in Appendix~\ref{appd}). 
We note that using \5 as a mass tracer of the outflowing gas implies several uncertainties since the derived outflow mass is sensitive to the ionization state and to the metallicity of the gas.
The latter cannot be directly measured from the optical line ratios of GMASS 0953 because of the presence of the AGN and we resort to using the mass-metallicity relation \citep{2016ApJ...827...74W} while the ionization state has been assumed (see Appendix~\ref{appd}).
On the other hand, hydrogen lines such as H$\beta$ and H$\alpha$ are better mass tracers because they do not depend on these quantities (see \citealt{2018NatAs...2..198H}). However, we do not find evidence of a blueshifted wing in the H$\beta$ profile probably because of the low S/N and the weakness of the emission. In the case of H$\alpha$ we highlighted in section~\ref{sec:kine} the presence of outflowing material possibly shown also by this line; however, the H$\alpha$ emission suffers from possible contamination of the \nii\ lines (see section~\ref{sub:kmos}) and from the AGN broad line region (see also \citealt{2015MNRAS.446.2394B}). 
For these reasons we do not use H$\alpha$ to derive the outflow properties.\par 
The outflow rate was estimated in two ways: with and without applying the dust extinction correction to the \5 luminosity.
As seen in section~\ref{sub:gnirs} GMASS 0953 is highly obscured, hence it seems reasonable to consider both scenarios.
To evaluate the intrinsic \5\ flux, we used the color excess $E(B-V)=0.8 \pm 0.3$ derived from the H$\alpha$ and H$\beta$ line ratio assuming a \citet{2000ApJ...533..682C} extinction curve.
The resulting ionized outflow mass $M_{\rm{of}}$ in the extinction corrected case amounts to $M_{\rm{of}} \simeq 1.2 \times 10^8 \ \text{M}_{\odot}$. Without the extinction correction this quantity drops to $\simeq 4.3 \times 10^6 \ \text{M}_{\odot}$.\par
 Once we have obtained the outflow mass, to evaluate the outflow rate it is necessary to assume a geometrical model to describe the wind. The low spatial resolution of our data does not provide information about the intrinsic geometry of the outflowing gas. For simplicity we used the model of \citet{2012A&A...537L...8C}, where the ionized gas is distributed in a conical region of radius $R_{\text{of}}$ and is expelled in a direction roughly perpendicular to the galaxy disc.
Under these assumptions, the outflow rate can be written as
\begin{equation}
\dot{M}_{\text{of}} = 3 M_{\text{of}} \frac{\vert v_{\text{of}}\vert}{R_{\text{of}}}
\label{eq:mdot2Corpo}
\end{equation}
where $v_{\rm{of}}$ is the outflow velocity estimated from the line fit of the GNIRS spectrum.
There is no general consensus on how to define outflow velocities (see \citealt{2018NatAs...2..198H}). In our work we followed the approach of \citet{2005ApJ...632..751R}, who defined the outflow maximum velocity as $v_{\text{of}} = {\rm FWHM}_{\rm broad}/2 + \vert \Delta v \vert$, where the first term is half the FWHM of the \5 blueshifted component and $\Delta v$ is the velocity offset between the \5 narrow and blueshifted component. The resulting velocity amounts to $v_{\text{of}} \sim 1200$ \kms . Using this $v_{\text{of}}$ value in  Eq.~\eqref{eq:mdot2Corpo} we are implicitly assuming that all the outflowing material is moving at this velocity. This could be possible due to projection effects, supported by the high inclination of the galaxy (see also \citealt{2012A&A...537L...8C, 2015MNRAS.446.2394B, 2015ApJ...799...82C}). 
We note that using a more conservative definition of $v_{\rm{of}}$, relying on the velocity offset $\vert \Delta v\vert \sim 530$ \kms , the outflow rate would be reduced at most by a factor $\sim 2$ and our results would not be considerably modified.
Finally, we used as the outflow radius $R_{\text{of}} $ half the FWHM of the emitting region visible in the SINFONI map, which amounts to $R_{\text{of}} \simeq$ 3.3 kpc (see also \citealt{2012MNRAS.426.1073H, 2014MNRAS.441.3306H}).
Substituting these values in Eq.~\eqref{eq:mdot2Corpo} and using the extinction corrected outflow mass, we found an ionized outflow rate $\dot{M}_{\text{of}} \sim 120$ M$_{\odot} \rm{yr}^{-1}$ that drops to $\sim 6$ M$_{\odot}$yr$^{-1}$ if the outflow is not extincted by dust.\\
\indent An indication of whether the outflow is affected or not by dust obscuration may come from the 1.4 mm continuum of GMASS 0953 \citep{2018MNRAS.476.3956T}, which is a robust tracer of dust.
After applying an astrometric correction (see Appendix~\ref{appa}), we overlapped the dust continuum map on the outflow emission detected with SINFONI (Figure~\ref{fig:alma}).
Interestingly, the dust and the \5 blueshifted component emit in the same region.
Therefore, it is plausible that the \5 luminosity may be dimmed by the dust along the line of sight, though it is also possible that the outflow may have already escaped the dusty region of the galaxy because of the high velocity.\par
Apart from the extinction correction and the adopted assumption about the outflow velocity, it should be noted that other uncertainties affect the outflow rate. 
The derivation of the outflow mass $M_{\rm{of}}$ that enters Eq.~\eqref{eq:mdot2Corpo} requires the knowledge of the electron number density of the gas (see Appendix~\ref{appd}). This quantity was derived from the \sii\ line ratio and amounts to 500 ${\rm cm}^{-3}$ (see section~\ref{sub:gnirs}). Unfortunately, the \sii\ emission is produced by the galaxy ISM and not by the wind since we did not find evidence of a broad component in the \sii\ line fit likely because of the low S/N.
Finally, we remind that we assumed a conical outflow for simplicity, but the true geometry is not known.
\subsubsection{Depletion timescale}
In order to properly gauge the impact of the AGN feedback 
on GMASS 0953 a comparison between the depletion timescale due to star formation with that associated to gas ejection is necessary. The former amounts to $\tau_{\rm{dep}}^{\rm SF} = M_{\rm{H}_2}/ \rm{SFR} \sim 150$ Myr \citep{2017A&A...602A..11P, 2018MNRAS.476.3956T}. 
The depletion timescale due to the outflow can be evaluated as the ratio between the mass of the ionized gas and the outflow rate, i.e. $\tau_{\rm{dep}}^{\rm OF, ion} = M_{\rm ion}/ \dot{M}_{\text{of}} \sim 2$ Myr (extinction corrected case). This very short value is however implausible if the outflowing material is supplied by the internal gas reservoir of the galaxy, which is much higher than $M_{\rm ion}$. Under the assumption that the outflow is depleting the whole ISM, we can write the depletion timescale as $\tau_{\rm{dep}}^{\rm OF, mol} \approx M_{\rm{H}_2}/ \dot{M}_{\text{of}}$, which is $\sim 270$ Myr using again the extinction corrected \5\ luminosity.  
We note that this value of $\tau_{\rm{dep}}^{\rm OF, mol}$ represents an estimate of the depletion timescale in the central region only, since the molecular gas is concentrated within a radius of 1 kpc \citep{2018MNRAS.476.3956T}.
From the comparison of the depletion timescales, it appears that the AGN-driven outflow and the star formation activity contribute in a similar way in depleting the gas reservoir of the galaxy.
We note however that the outflow would be much less effective in consuming the gas if the not-extincted value of the \5 luminosity were considered. Using the not-extincted luminosity we obtain indeed a depletion timescale $\tau_{\rm{dep}}^{\rm OF, mol} \sim 5.4$ Gyr, meaning that the AGN-driven outflow is negligible compared to the star formation activity.
Moreover, the latter may also drive galactic winds that would provide an additional contribution to gas consumption from star formation \citep{2017arXiv170109062H}.\\
\indent Apart from the extinction correction, another important uncertainty affects the outflow depletion timescale. 
We considered in the evaluation $\tau_{\rm{dep}}^{\rm OF, mol}$ the contribution of the ionized outflow only. Many studies show that this phase cannot be the dominant constituent of AGN outflows \citep{2017A&A...601A.143F}; in particular, molecular outflows were found to be up to two orders of magnitude  more massive than ionized outflows in active galaxies where both the phases have been observed \citep{2015A&A...580A.102C}.
As we mentioned in section~\ref{sub:ionized}, we did not find evidence of outflowing material in the molecular phase except for a feature of 2.5$\sigma$ blueshfted from the CO(6-5) main line.
The ALMA data can be used to set an upper limit to the molecular outflow rate of GMASS 0953.
We stacked the channels between $- 500$ \kms\ and $- 1200$ \kms\ , corresponding to the outflow emission in the SINFONI data (see Figure~\ref{fig:pv_iniz}, middle panel), in order to derive a 3$\sigma$ upper limit to the flux density of the CO(6-5) outflow ($S_{\rm CO}^{3\sigma}= 79\ {\rm mJy}$, assuming the outflow has the same size of the ionized one; see also \citealt{2018A&A...612A..29B}). This value was converted in ${\rm H}_2$ mass using the same assumptions of \citet{2018MNRAS.476.3956T}.
We hence derived an upper limit to the molecular outflow rate using Eq.~\eqref{eq:mdot2Corpo} and adopting the same velocity of the ionized outflow. The upper limit to the molecular outflow rate amounts to $\sim 2500\ {\rm M}_{\odot} {\rm yr}^{-1}$. This means that our observation does not allow us to detect outflow rates below this quantity, which can potentially contribute to gas consumption.\\ 
\indent  From the reported analysis we conclude that the role played by the AGN feedback in GMASS 0953 is quite unclear. 
However, despite the aforementioned uncertainties, the galaxy appears a promising candidate to quench the star formation activity in a total depletion timescale $\tau_{\rm dep}^{\rm TOT} = {\rm M}_{\rm H_2}/({\rm SFR} + \dot{M}_{\rm of}) \sim 10^8\ {\rm yr}$ (see also \citealt{2018A&A...612A..29B}) if the accretion from external gas is shutdown. This estimate is valid in both the extinction corrected/not-corrected case. 
\section{Conclusions}
\label{sec:conclusions}
In this work we studied the kinematics of the active galaxy GMASS 0953 from a multi-wavelength perspective. Our analysis was focused on the \5 and H$\alpha$ emission lines, which we inspected combining three spectroscopic dataset. 
The main conclusions of our work are summarized here:\\ 
\indent (i) We found evidence of multi-phase disc supported by the \5, H$\alpha$ and CO(J=6-5) emission lines, which have parallel velocity gradients extending in the same direction of the stellar component. We investigated the kinematics of \5 and H$\alpha$ independently using a three-dimensional modelling that corrects for beam smearing. The two lines present consistent rotation curves with an average rotation velocity $V_{\rm{ion}} = 203^{+17}_{-20}$ \kms\ at a distance of 13 kpc from the galaxy center.
As already known, the CO(J=6-5) traces rotation on a much smaller scale ($R \sim 1$ kpc) with a rotation velocity $V_{\rm{CO}} = 320^{+ 92}_{-53}$ \kms \citep{2018MNRAS.476.3956T}. 
Combining these two results, the arising picture is that GMASS 0953 may host a multi-phase disc with a rotation curve peaking in the central regions and a dynamical mass $> 1.3 \times 10^{11}$ M$_{\odot}$, consistent with the stellar mass.
The galaxy falls onto the 0.11 dex scatter of the $z = 0$ baryonic Tully-Fisher relation of \citet{2016ApJ...816L..14L}, suggesting that it is already kinematically relaxed.
On the other hand, the velocity dispersion of the disc of GMASS 0953 is poorly constrained by our kinematic model because of the low spatial resolution of the SINFONI and KMOS observations.\\ 
\indent (ii) Based on the analogies between the \5 and H$\alpha$ pv diagrams of GMASS 0953 and the local starburst M82, we also discussed the possibility of a galaxy-scale wind as an explanation of the observed gradient. Despite finding this scenario unlikely, we caution the reader over the changelling interpretation of gas velocity gradients in data with low spatial resolution. \\
\indent (iii) We discovered an ionized outflow extending on kpc-scale and likely associated to the nuclear activity of GMASS 0953. The outflow was highligted by a broad (FWHM $\sim $ 1400 \kms), blushifted wing in the \5 line profile with a velocity shift of $\Delta v = -535 \pm 152$ \kms\ from the systemic velocity. 
The blueshifted \5 emission was detected independently by the GNIRS and SINFONI spectrographs. 
Other hints of outflowing material in the ionized phase come from the position-velocity diagram of H$\alpha$ presented in this work (see also \citealt{2014ApJ...787...38F, 2014ApJ...796....7G}) and the UV absorption lines \citep{2013ApJ...779L..13C}.\\
\indent (iv) We derived the outflow rate of the ionized gas in order to assess the impact of the AGN feedback on the galaxy. We found that the outflow rate amounts to  $\sim 120$ M$_{\odot}$yr$^{-1}$, which drops to  $\sim 6$ M$_{\odot}$yr$^{-1}$ without applying the dust extinction correction.
We compared the gas depletion timescale due to the outflow with the one based on star formation finding that they are of the same order of magnitude (extinction corrected case). When both processes are considered we find a total depletion timescale of 10$^8$ Myr. This means that GMASS 0953 is expected to rapidly quench the star formation activity, if the accretion of external gas is shutdown.
However the role played by the AGN in this galaxy appears unclear due to the several uncertainties affecting the outflow rate. For example, if the outflow luminosity is not extincted by dust, the depletion timescale associated to the outflow drops to 5.4 Gyr, negligible compared to that due to star formation ($\tau_{\rm dep}^{\rm SF} \sim 150\ {\rm Myr}$). Besides, using our data we can only constrain the ionized phase while we are not accounting for possible gas expulsion in other phases (e.g. molecular).
Deeper ALMA observations will be crucial in this sense.\\
\indent Despite the remaining open questions, we stress the importance of multi-wavelength and multi-instrument studies of high-$z$ galaxies. They provide complementary details about the ISM that are essential to characterize the processes occurring in $z\sim 2$ massive systems.
Even though the uncertainties due to the observational limits of high-$z$ sources, multi-wavelength studies represent the most powerful approach to investigate the kinematics of $z\sim 2$ galaxies and the role played by AGN in their evolution.
\section*{Acknowledgements}
We gratefully thank N. M. F$\ddot{\rm{o}}$rster Schreiber and the KMOS$^{3\rm{D}}$ and SINS/zC-SINF team for sharing the reduced KMOS and SINFONI data analyzed in this work. We thank P. G. van Dokkum for the GNIRS spectrum and P. Rosati who provided the reduced MUSE data. We thank also P. Martini for having kindly made available the reduced data-cube of M82. 
AC, MT, FL acknowledge the support from the grants PRIN-MIUR 2015, PRIN 2017 and ASI n.2018-23-HH.0.
\bibliographystyle{mnras}
\bibliography{bibliografia} 
\appendix
\section{Astrometric corrections}
\label{appa}
We verified the astrometry of the SINFONI data against the \textit{HST} imaging of GMASS 0953.
In order to properly overlap the maps of \5 on the \textit{HST} images, we adjusted the data-cube coordinates imposing the matching between the peak of the SINFONI stacked continuum and the light peak determinated in the \textit{HST/H}-band image.
The coordinates of the peaks were both calculated through a two-dimensional Gaussian fit.
The correction resulted in a shift of + 1.26$''$ in R.A. and + 2.16$''$ in Dec. applied to the data-cube coordinates, after which we could nicely superimpose the SINFONI line maps on the \textit{HST} images (Figure~\ref{fig:HST_disc}).\\
\indent We verified also the astrometry of the ALMA data with respect to the \textit{HST} images of GMASS 0953.\
The systematic offset between ALMA and the \textit{HST} data in the GOODS-South is a well known problem that was underlined in many works (e. g. \citealt{2015MNRAS.452...54M}, \citealt{2017MNRAS.466..861D}).
We applied the correction of \citet{2018A&A...620A.152F} determined for the object AGS13: this acronym refers indeed to GMASS 0953, which was studied also in that work.
We shifted the coordinates of - 0.087$''$ in R.A. and + 0.291$"$ in Dec. In this way we could correctly overlap the dust continuum of the galaxy on the \textit{HST} images of GMASS 0953 (Figure~\ref{fig:alma}).
\section{KMOS line fitting}
\begin{figure*}
\begin{center}
\includegraphics[scale=0.7]{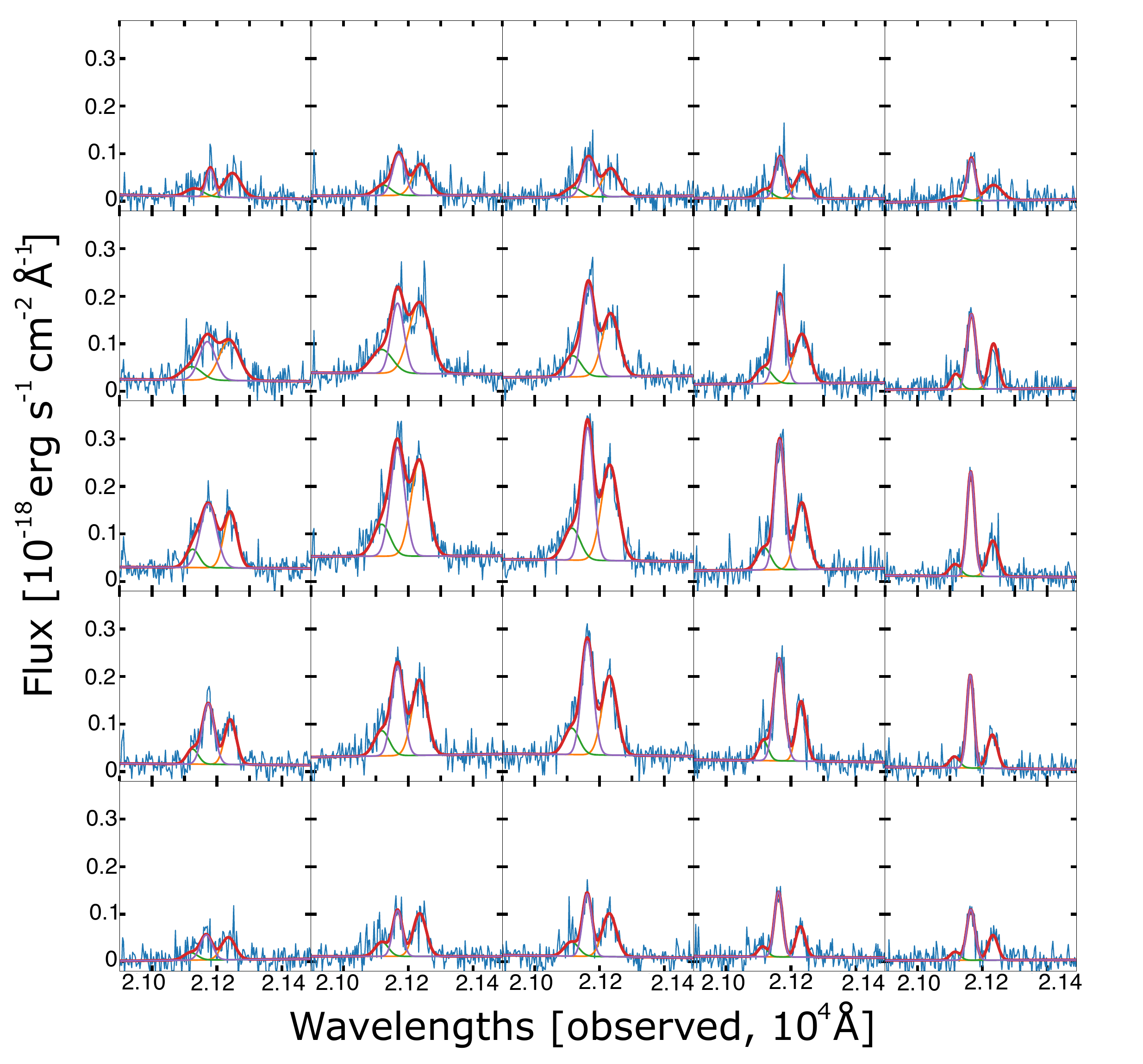}
\end{center}
\caption{Line fit of H$\alpha$ and \nii\ (KMOS data). Each spectrum was extracted from a pixel belonging to a section of $5 \times 5$ pixels, which corresponds to the central region of GMASS 0953. We modelled the data (blue line) with three Gaussian components and a linear continuum (colored lines). The centroids ratio of the \nii\ doublet and H$\alpha$ was fixed to the theoretical one. We imposed the same FWHM for the nitrogen lines and their flux ratio was fixed to 1:3. The total model is overlayed in red.}
\label{fig:fitKmos}
\end{figure*}

In order to study the kinematics of the H$\alpha$ line, it was necessary to separate its emission from the \nii\ doublet, which is blended with H$\alpha$, and to subtract the stellar continuum.
A line fit was hence performed in each pixel of the KMOS data-cube containing the stellar and \nii\ emission (Figure~\ref{fig:fitKmos}).
We selected a region of 5 $\times$ 5 pixels corresponding to the galaxy central part
because the nitrogen and the stellar emission were mainly confined in a region of this size ($\sim 6.7$ kpc).
For each pixel we extracted a spectrum
that was modelled with three Gaussians and a linear continuum. The centroids ratio of the \nii\ doublet and H$\alpha$ was fixed to the theoretical one (0.995 for the \nii\ lines and 0.998 for the [N II]$\lambda$6548 and H$\alpha$ ratio); we also imposed the same FWHM for the nitrogen lines and fixed their intensity ratio to 1:3\footnote{We executed also a fit in which the intensity ratio of \nii\ was left free, but it showed no significant differences.}.
We then subtracted from each pixel in the data-cube the corresponding \nii\ models and the continuum level. In this way we obtained a data-cube that contains only the H$\alpha$ emission, which was used to study the kinematics.
\label{appb}
\section{Inclination angle}
\label{appc}
\begin{figure}
\begin{center}
\includegraphics[scale=0.6]{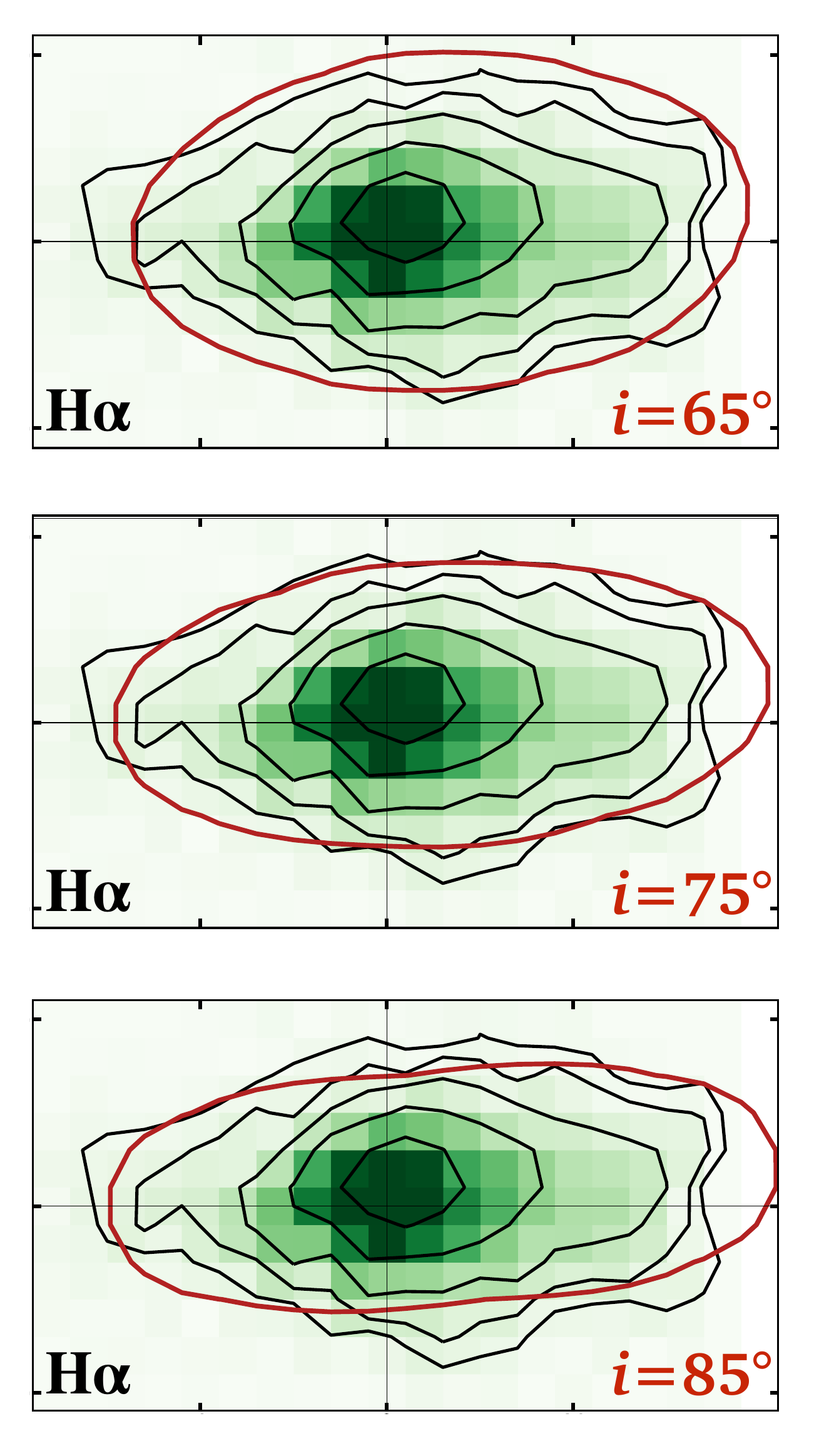}
\end{center}
\caption{Comparison between the H$\alpha$ flux map (black contour and colors) and the model maps (red contour, 3$\sigma$ level) obtained using an azimuthally averaged H$\alpha$ profile for fixed inclination angles. The inclination angle was varied from $i = 60$\degr\  to $i = 85$\degr . We report here the models corresponding to  $i = 65$\degr\ and  $i = 85$\degr\ (top and bottom panel)
and the one with $i = 75$\degr\ (middle panel), which we fixed as the disc inclination. 
The more plausible inclination angles are included in the range $70$\degr\ $< i < 80$\degr\ while for $i < 70$\degr\ and $i > 80$\degr\  the model varies significantly from the data.}
\label{fig:inciinHa}
\end{figure}

The disc inclination in the $^{3\rm{D}}\rm{\textsc{Barolo}}$ model was determined by comparing the shape of the H$\alpha$ total flux map with synthetic maps obtained for fixed inclination angles (from 60\degr\ to 85\degr) using an azimuthally averaged H$\alpha$ surface brightness profile (Figure~\ref{fig:inciinHa}).
We used the H$\alpha$ line instead of \5 because of the higher flux of the receding half, which allows to better examine the geometry of the disc. This (by eye) comparison allowed us to exclude inclination angles lower than 70\degr\ and greater than 80\degr\ and as a fiducial average value we took $i = 75$\degr\ for all the rings. The reliability of our choice was confirmed by the residuals between the data and the model evaluated for different inclination angles, which show a minimum corresponding to $i = 75$\degr .
Therefore we fixed $i = 75$\degr\ for both the \5 and H$\alpha$ disc models. We considered the uncertainty due to the inclination angle in the range 70\degr\ $< i <$ 80\degr\ when we calculated the errorbars of the \5 and H$\alpha$ rotation velocity. However, given the high value of $i$, the uncertainty on the inclination does not have a big impact on the derived rotation velocity.
\section{Outflow mass and outflow rate}
\label{appd}
To evaluate the outflow mass and the outflow rate traced by \5 we followed the approach of \citet{2012A&A...537L...8C} and \citet{2015A&A...580A.102C}. 
We started evaluating the outflow mass, i.e. the total mass of the ionized gas expelled, from the luminosity of the \5 blueshifted component.
The luminosity of the [O III] $\lambda$5007 can be written as
\begin{equation}
L_{\text{[O III]}} = \epsilon_{\lambda 5007} f V_{\text{of}}
\label{eq:L1}
\end{equation}
where $\epsilon_{\lambda 5007}$ is the line emissivity, $V_{\text{of}}$ the total volume occupied by the ionized outflow and $f$ is the filling factor, i. e. the ratio between the volume of the [O III] $\lambda$5007 emitting gas and the total volume: we worked indeed under the hypotesis that [O III] is produced by a fraction of the outflowing ionized gas. We used the emissivity value reported by \citet{2012A&A...537L...8C} for a temperature $T$ = 10$^4$ K:
\begin{equation}
\epsilon_{\lambda 5007}=1.11 \times 10^{-9} h \nu_{\lambda 5007} n_{\text{O}^{2+}} n_\text{e}\quad \text{[ erg s$^{-1}$cm$^{-3}$ ]}
\label{eq:emiss}
\end{equation}
where $n_{\text{O}^{2+}}$ is the number density of the doubly ionized oxygen and $n_\text{e}$ the electron number density. 
Under the reasonable assumption that all the oxygen is in the ionized form, we can espress $n_{\text{O}^{2+}}$ as
\begin{equation}
n_{\text{O}^{2+}}= \frac{n_\text{O}}{n_{\text{H}}} \frac{n_{\rm{H}}}{n_\text{e}} n_\text{e}
\label{eq:no}
\end{equation}
where $n_\text{O}$ and $n_{\text{H}}$ are the neutral oxygen and hydrogen number densities. The electron number density $n_\text{e}$ can be written as $n_\text{e} \approx n_{\text{H}^{+}}+ 1.5 n_{\text{He}}$ for a helium mixture made up by equal parts of He$^{+}$ and He$^{++}$ \citep{2006agna.book.....O}. Since the number density of the ionized helium $n_{\text{He}}$ is in good approximation 0.1$n_{\text{H}^{+}}$, under the hypothesis that $n_\text{H} \approx n_{\text{H}^+}$, Eq.~\eqref{eq:no} becomes
\begin{equation}
n_{\text{O}^{2+}}=\frac{n_\text{O}}{n_{\text{H}}} (1.15)^{-1} n_\text{e} = 10^{\log\left(\frac{n_\text{O}}{n_{\text{H}}}\right)_{\odot}}10^{\left[ \frac{O}{H}\right]} (1.15)^{-1} n_\text{e}
\label{eq:no1}
\end{equation}
where $\left[ \frac{O}{H}\right]\equiv \log\left(\frac{n_\text{O}}{n_{\text{H}}}\right) - \log \left(\frac{n_\text{O}}{n_{\text{H}}}\right)_{\odot}$ is the metallicity.
Substituting Eq.~\eqref{eq:no1} in Eq.~\eqref{eq:emiss} and Eq.~\eqref{eq:L1} we obtain
\begin{equation}
L_{\text{[O III]}}=4.727 \times 10^{-13} h \nu_{\lambda 5007} n_\text{e}^2 10^{\left[ \frac{O}{H}\right]} f V_{\text{of}}\quad \text{[ erg s$^{-1}$ ]}
\label{eq:Lion}
\end{equation}
We used for the oxygen abudance of the sun $(n_{\rm{O}}/n_{\rm{H}})_{\odot}$ the value of \citet{2009ARA&A..47..481A}.
If we neglect the elements heavier than helium, the ionized gas mass $M_{\text{of}}$ of the [O III] $\lambda$5007 emitting clouds is essentially given by
\begin{equation}
M_{\text{of}} = (n_{\text{H}^{+}} m_\text{p} + n_{\text{He}}m_{\text{He}})f V_{\text{of}}
\label{eq:mion}
\end{equation}
where $m_\text{p}$ and $m_{\text{He}}$ are the proton and helium mass. 
Since $m_{\text{He}} \approx 2m_\text{p}$ and expressing $n_{\text{H}^{+}}$ and $n_{\text{He}}$ in terms of $n_\text{e}$, Eq.~\eqref{eq:mion} becomes
\begin{equation}
M_{\text{of}} = 1.04\ n_\text{e} m_\text{p} f V_{\text{of}}
\label{eq:Mion}
\end{equation} 
which combined with Eq.~\eqref{eq:Lion} gives
\begin{equation}
M_{\text{of}} = \frac{1.04\quad m_\text{p}}{4.727 \times 10^{-13} h \nu_{\lambda 5007} n_\text{e} 10^{\left[ \frac{O}{H}\right]}}L_{\text{[O III]}} \quad \text{[ g ]}
\label{eq:m_of}
\end{equation}
We measured the electron number density $n_\text{e}$ from the \sii\ line ratio (see section~\ref{sub:gnirs}). The metallicity was derived using the mass-metallicity relation of \citet{2016ApJ...827...74W} and amounts to $\left[ \frac{O}{H}\right] = \log(0.9)$ for a stellar mass of 10$^{11}$ M$_{\odot}$.\\
\indent Once we have the outflow mass we can evaluate the outflow rate, i.e. the ionized gas mass expelled per unit time. In order to calculate it, we adopted a simple geometrical model to describe the outflow, as done by \citet{2012A&A...537L...8C}, where the material is distributed in a conical region with an opening angle $\Omega$ and radius $R_{\text{of}}$ and is expelled in a direction almost perpendicular to the galaxy disc. 
Under this assumption, the outflow rate can be written as
\begin{equation}
\dot{M}_{\text{of}} = \rho v_{\text{of}} \Omega R_{\text{of}}^2
\label{eq:mdot}
\end{equation}
where $\rho$ is the mass density of the ionized gas and $v_{\text{of}}$ the outflow velocity. Assuming that the ejected material is located in a volume $V_{\text{of}}=\frac{4}{3} \pi R^3_{\text{of}} \frac{\Omega}{4\pi}$ and simply writing $\rho$ as $\rho = M_{\text{of}}/V_{\text{of}}$, Eq. ~\eqref{eq:mdot} becomes
\begin{equation}
\dot{M}_{\text{of}} = 3 M_{\text{of}} \frac{v_{\text{of}}}{R_{\text{of}}}
\label{eq:mdot2}
\end{equation}
by which it is possible to estimate the outflow rate of the ionized gas.
\\
\bsp	
\label{lastpage}
\end{document}